\documentclass[11pt,a4paper,twoside]{article}
\pdfoutput=1
\usepackage[T1]{fontenc}
\usepackage[utf8]{inputenc}
\usepackage{cite}
\usepackage{fancyhdr}
\usepackage{varwidth}
\usepackage{setspace}
\usepackage[hyperindex,breaklinks]{hyperref}

\usepackage{changepage}

\usepackage[color=red,linecolor=red,bordercolor=red,backgroundcolor=none,textsize=footnotesize,textwidth=2.35cm]{todonotes}

\usepackage{enumitem}

\usepackage{amsmath,amsfonts,amssymb,mathtools}
\usepackage{bbm}
\usepackage[low-sup]{subdepth}

\usepackage[format=plain,indention=2em,labelfont=bf,font=small,labelsep=period]{caption}

\newcommand{\nn}{\nonumber\\}

\DeclareRobustCommand{\bfrac}[2]{%
	\mathchoice{\frac{\raisebox{-0.5ex}{$#1$}}{\raisebox{0.ex}{$#2$}}}%
	{\frac{\raisebox{0.ex}{$\scriptstyle#1$}}{\raisebox{0.1ex}{$\scriptstyle#2$}}}%
	{\frac{#1}{#2}}%
	{\frac{#1}{#2}}%
}

\newcommand{\intd}[2][]{\int_{#1}\,\mathrm{d}^{#2}\!x\ }

\def\p{\raisebox{-1pt}{$'$}}

\def\tilde{\widetilde}

\def\Langle{\big\langle}
\def\Rangle{\big\rangle_0}

\def\de{\!\mathrm{d}}

\def\g{\sqrt{\!\smash[t]{-g}\phantom{|}}}
\def\gh{\sqrt{\!\smash[t]{-\hat{g}}\phantom{|}}}
\def\nud{\sqrt{\smash[t]{\bfrac{d^2}{4\,}+m_\varphi^2}\,\vphantom{\frac{1}{4}}}}

\def\phiz{\phi_0}
\def\tphiz{\tilde{\phi}_0}
\def\rhoz{\rho_0}
\def\trhoz{\tilde{\rho}_0}
\def\piz{\pi_0}
\def\tpiz{\tilde{\pi}_0}

\def\vth{\vartheta}

\def\sfI{\mathsf{I}}
\def\sfK{\mathsf{K}}

\def\CO{\mathcal O}
\def\CL{\mathcal L}
\def\ImO{\mathrm{Im}\mathcal{O}}

\def\ReO{\mathrm{Re}\mathcal{O}}

\def\vardel{{\delta\!\!\phantom{.}}}

\def\mod{{\hspace{.5pt}\rule[-0.15mm]{0.75pt}{3.8mm}\hspace{.5pt}\vec{k}\hspace{1pt}\rule[-0.15mm]{0.75pt}{3.8mm}\,}}
\newcommand{\modq}[1]{{\hspace{.5pt}\rule[-0.15mm]{0.75pt}{3.8mm}\hspace{.5pt}\vec{k}\hspace{1pt}\rule[-0.15mm]{0.75pt}{3.8mm}\hspace{1pt}\raisebox{6.5pt}{$\scriptstyle #1$}}}
\def\veckk{\vec{k}\hspace{1pt}\raisebox{5.7pt}{$\scriptstyle2$}}

\usepackage[left=2.8cm,right=2.8cm,top=2.5cm,bottom=3cm]{geometry}
\numberwithin{equation}{section}
\begin{document}	
\begin{titlepage}
\thispagestyle{empty}
\topmargin=3.8cm
\oddsidemargin=10pt
\begin{center}
	\begin{varwidth}[t]{1.1\textwidth}
	\centering\setstretch{2.5}
	\textsc{\huge
		The Ward identity of Symmetry Breaking}\\
	\textsc{\huge
		from Holographic Renormalization}\\[11pt]
	\textsc{\Large
		a practical handbook}
	\end{varwidth}
		
		\vspace{6em}
		
		{\large Andrea Marzolla$^{\S}$}
		
		\vspace{1em}
		\parbox[t]{0.7\textwidth}{\itshape \footnotesize \centering
			{$^{\S}$Physique Th\'eorique et Math\'ematique and International Solvay Institutes, Universit\'e Libre de Bruxelles, C.P. 231, 1050 Brussels, Belgium.}
			}

	\vspace{5.5cm}

{\bf \normalsize \textsc{Abstract}}
\vskip 1em
	\begin{minipage}[t]{\textwidth}
	\noindent
	Notes on holographic renormalization as a tool for UV analysis and derivation of the two-point Ward identity encoding symmetry breaking. Goldstone theorem is reviewed, discussing in particular the modifications required by its extension to non-relativistic field theories. The Ward identity encoding both spontaneous and explicit breaking are derived by general principles. Holographic renormalization is presented in a quite operational fashion. The derivation of the Ward identity in holography is illustrated by an example.
	\end{minipage}
\end{center}

\end{titlepage}

\clearpage{\pagestyle{empty}\cleardoublepage}
\pagenumbering{roman}

\subsection*{Disclaimer for the reader}

These notes represents an extract of the author's PhD thesis ``Symmetry breaking and Goldstone theorem in holographic strongly coupled field theories -- Relativistic and non-relativistic examples'', defended on September 29, 2017, at the \emph{Universit\'e Libre de Bruxelles}. For this unique reason the spirit and most of the provided examples are taken from or closely related to the works of the author. 

The full text of the thesis is available at the link here below.

\vskip 1em
\noindent\href{http://difusion.ulb.ac.be/vufind/Record/ULB-DIPOT:oai:dipot.ulb.ac.be:2013/258338/Holdings}{\sf%
http://difusion.ulb.ac.be/vufind/Record/ULB-DIPOT:oai:dipot.ulb.ac.be:2013/258338/Holdings}

\clearpage{\pagestyle{empty}\cleardoublepage}
\pagenumbering{arabic}
\section*{Introduction}
\markboth{Introduction}{}
\phantomsection

Spontaneous symmetry breaking, with its associated Goldstone massless modes, is an ubiquitous phenomenon in physics, especially in particle physics and condensed matter physics. Emblematically, its first observation was at the intersection of the two fields~\cite{Nambu:1960tm}. It finds its theoretical grounds in the Goldstone theorem~\cite{Goldstone:1961eq}, which is a very well defined mathematical statement, at least in the relativistic framework where it has been first proven~\cite{Goldstone:1962es}. 

If relativistic invariance is an essential ingredient in particle high energy physics, it is likely disregarded in condensed matter systems, where Goldstone theorem has known many successful applications in describing phase transitions (superconductivity, Bose-Einstein condensation, etc). So, the extension of Goldstone theorem to non-relativistic field theories~\cite{Klein:1964ix,Lange:1965zz,Lange:1966zz,Guralnik:1967zz} has drawn attention since the early stages of its conception. In the absence of Lorentz invariance, the physics of Goldstone bosons (GB) gets richer, and its mathematical foundations have to be revised and extended.

For instance, the one-to-one matching between broken symmetry generators and Goldstone massless modes breaks down in non-relativistic settings~\cite{Nielsen:1975hm}. Moreover, the dispersion relations of gap-less modes can have more diversified forms than the relativistic one~$\omega=\mod$. We can have Goldstone bosons with linear dispersion relation,~$\omega=c_s\mod$, but with velocity now non-necessarily equal to the speed of light~$c\equiv1$. But we can also have Goldstone modes which present dispersion relations with quadratic or higher-order low-momentum behavior. In particular, in non-relativistic theories with spontaneously broken \emph{non-abelian} global symmetries, the commutator of two broken charges may get a non-vanishing expectation value. In such situation, the two corresponding broken generators give rise to a single Goldstone mode, which presents in turn a quadratic dispersion relation,~$\omega\propto\modq{2}$~\cite{Brauner:2007uw,Watanabe:2011ec,Watanabe:2012hr}. The second `would-be massless' mode typically gets a mass proportional to the Lorentz breaking parameter~\cite{Kapustin:2012cr}. 

In more recent years, another input for theoretical physicists comes from condensed matter systems: a large variety of experimental devices presents `strange'  behaviors, which can be ascribed to strong coupling effects. By strange it is meant that are not predicted by present-day available theoretical models (weakly-coupled Fermi-liquid theory), which are based on perturbation theory, and so fail at strong coupling.
 
Of course, whatever is based on symmetries is non perturbative. So, the physics of symmetry breaking, spontaneous or (slightly) explicit, is particularly relevant for strongly coupled systems, where it can constitute some of the (few) things we are able to say about. To this extent, a renewed interest in such an old and broadly discussed topic, enlarged towards these unconventional non-relativistic frameworks, may be justified.

At the same time, new techniques, that be capable to deal with strong coupling, are called for. Holography, which is a quite recent and active field of research, is a promising candidate. Since it has been conjectured, exactly twenty years ago~\cite{Maldacena:1997re}, as a duality between compactifications of string theory to Anti de Sitter spacetime (AdS), in the classical limit, and supersymmetric conformal field theories on the boundary of AdS, in the 't Hooft limit (large~$N$, strong coupling), the holographic principle has been extended to a much larger class of dualities. In such dualities the classical limit of the gravitational theory in the bulk corresponds to the strong coupling limit of the quantum field theory on the boundary. 

So, holography potentially constitutes a powerful tool to investigate the strong coupling regime of field theory. Indeed, the classical regime of gravity, which is generically well understood and under control, can be used to gain insights over the physics of field theory at strong coupling (confinement, bound states, quantum phase transitions, etc).

Non-relativistic behaviors are observed in many of the above-mentioned condensed matter experiments (high temperature superconductors, cold atoms at unitarity, etc). So, the ambition to apply holography to condensed matter has raised the challenge of extending the correspondence to non-relativistic backgrounds~\cite{Son:2008ye,Balasubramanian:2008dm,Kachru:2008yh,Taylor:2008tg,Guica:2010sw,Taylor:2015glc}.
%
%

The main focus of these notes will be the precise relations (Ward identities) that occur in case of spontaneous or explicit symmetry breaking, among two-point correlation functions of a conserved current and the scalar operator which is charged under it. These Ward identities, which are derived in full generality in field theory, can be retrieved in holography through a pure boundary analysis\footnote{
	Since we are interested in two-point functions of scalars and conserved currents, gravity is kept non-dynamical throughout all this work. The completely boundary analysis allow to neglect the back-reaction of other fields on the metric, and treat the metric as a fixed background.
}, constituting thus an independent check of the correspondence. Such boundary analysis typically requires the treatment of bulk terms that diverge at the boundary, which can be consistently removed by a procedure called holographic renormalization~\cite{deHaro:2000vlm,Bianchi:2001de,Bianchi:2001kw}.

The holographic renormalization procedure depends sensitively on the number of space-time dimensions, on the scaling dimensions of the considered field, on the background metric, etc. So, it can in some occasions present subtleties, which are addressed in this notes. 

\medskip

So, we will start with reviewing in Section~\ref{symmbreakQFT}, in a pure field theoretical approach, the physics of symmetry breaking. In Section~\ref{GoldstoneSSB} we recall the Noether theorem and the relation between symmetry and conserved currents, we introduce the spontaneous symmetry breaking mechanism, and we reproduce the original non-perturbative proof of Goldstone theorem~\cite{Goldstone:1962es}. In this section we are close to the approach of Weinberg's textbook~\cite{Weinberg:1996kr}. In Section~\ref{ColemanTh} we discuss the particular case of 1+1 dimension and the no-Goldstone theorem by Coleman~\cite{Coleman:1973ci}, commenting in particular on its quantum nature~\cite{Ma:1974tp}, and on the peculiarities of the large $N$ limit~\cite{Witten:1978qu}. 

Then, in Section~\ref{explicitWI}, we allow for explicit breaking on top of spontaneous one, and we derive in maximal generality (without specializing to any specific theory) the corresponding Ward identity, which in the purely spontaneous case reveal the presence of a massless mode, and in the case of (little) explicit breaking imply the notorious GMOR relation~\cite{GellMann:1968rz} for the (little) mass of the pseudo-Goldstone boson. Our derivation is standard, but partially novel in its formulation, which follows the one presented in~\cite{Argurio:2015wgr}. For these initial sections, we also found a helpful tool in~\cite{Soldati1}.

Then, in Section~\ref{nonrelGoldstoneTh}, we abandon Lorentz invariance, and we discuss the issues of the generalization of Goldstone theorem to non-relativistic field theories. We considerably owe to~\cite{Guralnik:1967zz,Brauner:2010wm} in the presentation, nonetheless we give a different, original derivation of non-relativistic dispersion relations, starting from the symmetry breaking Ward identity.

\medskip

The second section is dedicated to the holographic correspondence~\cite{Maldacena:1997re,Witten:1998qj}, and in particular to the technique of holographic renormalization~\cite{deHaro:2000vlm,Bianchi:2001de,Bianchi:2001kw}, which is the main computational tool presented here. After giving a sketchy review of the holographic principle, close in spirit to~\cite{Polchinski:2010hw}, in Section~\ref{HoloRen} we introduce the renormalization procedure by a simple, but rather complete example, presented in a quite personal and practical way.

\medskip

We conclude with an example, in Section~\ref{HoloPGB}, where the physical elements and the computational techniques are illustrated by a simple, but paradigmatic holographic model~\cite{Argurio:2015wgr}. The Ward identities for concomitant spontaneous and explicit symmetry breaking, shown in Section~\ref{explicitWI}, are retrieved holographically, and the presence of the (pseudo-)Goldstone boson in the dual theory is unveiled. The reader is finally addressed towards further examples in the literature, which consider non-relativistic or lower-dimensional setups.


\noindent\hrulefill
\vspace*{-12.5mm}
\tableofcontents
\vspace*{1mm}
\noindent\hrulefill

\vspace*{2em}

\subsection*{Notations and conventions}
\markboth{Notations and conventions}{Notations and conventions}
\phantomsection
\addcontentsline{toc}{section}{Notations and conventions}

We briefly establish here the main notations and conventions that are used throughout this manuscript. First of all, we work in natural units~$c=1=\hslash$, and in mostly-plus Lorentzian signature. Then, for the majority of prescriptions regarding analytic continuation and $\pm i$ factors in quantum field theory prescriptions, we employ the same conventions as Weinberg in his textbooks~\cite{Weinberg:1995mt,Weinberg:1996kr}. Einstein convention about summation over repeated indices is also understood, unless otherwise specified.

In the holographic sections, we use roman indices~$m,n,r,s$ for the~\mbox{$(d\!+\!1)$-}di\-men\-sion\-al bulk coordinates, saving greek letters~$\mu,\nu,\rho,\sigma$ as indices of the flat Minkowski space on the boundary. When, for non-relativistic examples, a distinction between temporal and spatial components is required, we make use of roman characters~$i,j,k,l$ for the spatial components. The notation~$x^\mu=(t,\vec{x})$ is also widely employed. So, for instance, for the AdS$_{d+1}$ metric in the Poincar\'e patch, we will write
\begin{equation*}
\de{s^2} = \de{x_m}\de{x^m} = r^{-2}\left(\de{r^2}+\de{x_\mu}\,\de{x^\mu}\right) = r^{-2}\left(\de{r^2}-\de{t^2}+\de{x_i}\,\de{x_i}\vphantom{\Big|}\right)\ .
\end{equation*}

Moreover, in the holographic parts we will specify the space-time dimensions in the integration measure, in order to allow to distinguish bulk terms from boundary terms. Otherwise, by simply~{ $\de{x}$} we will indicate integration over all space-time coordinates, and by~{ $\de\vec{x}$} integration over all spatial coordinates.

Finally our conventions for Fourier transformations to momentum space are
\begin{equation*}
f(x) = \int\!\de{k}\ e^{ikx} f(k) = \int\!\de{\omega}\:\de\vec{k}\ e^{-i\omega t+ik_ix_i} f\big(\omega,\,\vec{k}\big)\ ,
\end{equation*}
so that we schematically have $\partial_t\to-i\omega$, and~$\partial_i\to+ik_i$.

\clearpage{\pagestyle{empty}\cleardoublepage}
\section{Symmetry breaking in quantum field theory}	\label{symmbreakQFT}

In this section we want to review some very well known field-theoretical results, re-deriving them in a perspective that will make them easier to be compared to the corresponding results subsequently obtained through holographic techniques.

So we will put on the table the main physical ingredients of these notes: internal symmetries and Noether conserved currents, spontaneous breaking and Goldstone theorem, explicit breaking and Ward identities.

\subsection[Spontaneous breaking and Goldstone theorem]{Spontaneous breaking and Goldstone\\theorem}
\label{GoldstoneSSB}

We consider an action functional~$S$ on a $d$-dimensional space-time, depending on some field content $\{X_A\}$, and derivatives. The fields~$X_A$ represent all the real scalar degrees of freedom occurring in the action. We assume that such action is invariant under some internal (\emph{i.e.}~spacetime independent) continuous symmetry:
\begin{equation}	\label{invS}
\vardel_\alpha S\big[\big\{X_A,\partial_\mu X_A\big\}\big] = 0\ ,	\qquad	\text{for }\	\vardel_\alpha X_A=T^a_{AB}X_B \alpha^a\ ,
\end{equation}
where repeated indices imply a summation, and $\alpha^a$ is the infinitesimal parameter associated to the transformation performed by the generator~$T^a$ of the considered symmetry group, in the representation under which the field~$X_A$ transforms.

Then we can apply this variation on the Lagrangian density, obtaining
\begin{align}
\vardel_\alpha S = \intd{}\, \vardel_\alpha\CL &= 
	\intd{}\, \bigg[ \bfrac{\delta\,\CL}{\delta{X_A}}\,\vardel_\alpha X_A +\bfrac{\delta\,\CL\quad}{\delta\partial_\mu{X_A}}\,\vardel_\alpha\partial_\mu{X_A} \bigg] \nn
&=
	\intd{}\, \bigg[ \bigg(\bfrac{\delta\,\CL}{\delta{X_A}} -\partial_\mu\bfrac{\delta\,\CL\quad}{\delta\partial_\mu{X_A}}\bigg)\: \vardel_\alpha X_A +\partial_\mu\bigg(\bfrac{\delta\,\CL\quad}{\delta\partial_\mu{X_A}}\, \vardel_\alpha X_A\bigg) \bigg]	\nn
&=
	\intd{}\, \partial_\mu\bigg(\bfrac{\delta\,\CL\quad}{\delta\partial_\mu{X_A}}\, T^a_{AB}X_B\bigg)\; \alpha^a\ , 	\label{divJvanish}
\end{align}
where we have integrated by part, and used the equations of motion for the classical fields~$X_A$. We can define a current
\begin{equation}	\label{conscurr}
{J^a}^\mu = -\bfrac{\delta\,\CL\quad}{\delta\partial_\mu{X_A}}\, T^a_{AB}X_B\ 
\end{equation}
for each generator of the symmetry group. Then, the invariance of the action under the considered symmetry~\eqref{invS} implies that the integrated expression~\eqref{divJvanish} has to vanish. If we assume that the spatial part of the current vanishes at large enough distances, which is always true for sufficiently causal theories, then we have
\begin{align}
0= \intd{}\, \partial_\mu {J^a}^\mu &= 
	\int_{0}^{T}\!\de{t}\int_{\mathbb{R}_{d-1}\!}\!\de{\vec{x}}\, \big(-\partial_tJ^a_t+\partial_i^{\vphantom{a}}J^a_i\big) 	= 		\label{divcurrboundary}\\
&=
	-\int_{\mathbb{R}_{d-1}\!}\!\de{\vec{x}}\ J^a_t(t,\vec{x})\Big|_0^T = Q^a(0)-Q^a(T)\ , \nonumber
\end{align}
where 
\begin{equation}	\label{conscharge}
Q^a = \int\!\de\vec{x}\ J^a_t(t,\vec{x}) =
	\bfrac{\delta\,\CL\;\;}{\delta\partial_t{X_A}}\, T^a_{AB}X_B \equiv 
		\int\!\de\vec{x}\ \Pi_A T^a_{AB} X_B\ ,
\end{equation}
with~$\Pi_A$ being the canonical conjugate momenta of the field~$X_A$.

We have thus obtained the central result of Noether theorem: for any given symmetry of the action a quantity conserved in time arises, the conserved charge. However, the fact that the integral~\eqref{divcurrboundary} vanishes does not imply that the current~\eqref{conscurr} is locally conserved. It can still be~$\partial_{\mu}{J^a}^\mu=\partial_{\mu}\Omega^\mu$, with~$\Omega^\mu$ vanishing at infinity. Then we can redefine the current as ${\tilde{J}{}^a}^\mu={J^a}^\mu-\Omega^\mu$, so that~$\partial_\mu{\tilde{J}{}^a}^\mu=0$. Furthermore, the current is still defined up to en equivalence class defined by
\begin{equation*}
J^{a\mu} \;\longrightarrow\; J^{a\mu} +\partial_{\nu}K^{[\mu\nu]}\ ,
\end{equation*}
where the divergence of any anti-symmetric tensor is annihilated by construction under a further action of~$\partial_{\mu}$. So, the conserved current is actually the part which is not trivially divergence-less\footnote{
	In the language of differential geometry the current is a closed form which is not exact.}
Anyway, the integration~\eqref{divcurrboundary} ensures that the conserved charge~\eqref{conscharge} is unambiguously defined.

Let us now consider the very common case where the action is invariant under a certain symmetry, whereas the ground state is not. Indeed, when a system possesses a certain continuous global invariance, the space of vacua likely does as well: the lowest energy state is degenerate, and gives a continuum of possible vacua, related to each other by transformations of the symmetry group.\footnote{%
	The idea holds equivalently for discrete symmetries, which in turns gives a discrete sets of vacua. However, the follow-up of the discussion and Goldstone theorem are only valid for continuous symmetries (see for instance the example of the real scalar field in Goldstone's original paper~\cite{Goldstone:1961eq}).}
Then, the ground state of the system is just one specific vacuum out of the continuum of possible ones, which thus breaks the invariance. This phenomenon is called \emph{spontaneous} symmetry breaking, as opposed to \emph{explicit} symmetry breaking, where instead the symmetry is broken at the level of the action and the current is not conserved anymore.

The discussion here is completely general, but if we wish, we can just think, as an example, of the paradigmatic toy-model of a complex scalar with Goldstone Mexican-hat potential~\cite{Goldstone:1961eq}. In that case the action is invariant under a global U$(1)$, and the potential yields a continuum of vacua, arranged on a circle in the phase space, covered by the U$(1)$ transformations. A specific choice of vacuum on the ground circle will break the U$(1)$ invariance.

We are now ready to see the physical consequences of such situation, which are condensed in the statement of the celebrated Goldstone theorem~\cite{Goldstone:1961eq}: for each spontaneously broken generator of a continuous global internal symmetry, a zero-energy mode appears in the spectrum. This statement strictly holds in the context of relativistic invariance and internal symmetries, and so will the proof we are going to report here after. Nonetheless, the statement can be relaxed and generalized to the non relativistic case, as we will see in the following. Goldstone theorem can also be applied, in some cases, to the spontaneous breaking of space-time symmetries~\cite{Low:2001bw,Watanabe:2013iia,Hayata:2013vfa,Brauner:2014aha}, which however will not be considered in these notes.

\subsubsection[Relativistic Goldstone theorem]{General proof of relativistic Goldstone theorem}

We now outline the general proof (not relying on perturbation theory) of Goldstone theorem, first given in~\cite{Goldstone:1962es}, and presented in more detail in chapter~19 of Weinberg's textbook~\cite{Weinberg:1996kr}. Our assumptions are: (1) the invariance under a global continuous symmetry, which gives a conserved current and the associated conserved charge; (2) the existence of a vacuum expectation value (vev) which is not invariant under that symmetry, thus spontaneously breaking the symmetry. Translational invariance of the vacuum, and Lorentz invariance are also ingredients that we will make use of in the proof.

Using the defining expressions for conserved current~\eqref{conscurr} and charge~\eqref{conscharge}, and the canonical commutation relation for quantum fields, that is
\begin{equation}
\big[X_A(t,\vec{x}),\Pi_A(t,\vec{y})\big]=i\delta(\vec{x}-\vec{y})\ ,
\end{equation}
we can restate the spontaneous symmetry breaking condition as follows
\begin{equation}	\label{ssbcomm}
\Langle\big[Q^a,X_A(0)\big]\Rangle = -i\,T^a_{AB}\Langle X_B(0)\Rangle \neq 0\ , \qquad
	\text{for some }\, A\ .
\end{equation}
Notice that this definition is equivalent to saying that the vacuum be not invariant under the symmetry generator; if it was the case, then~$Q^a|0\rangle=0$, and consequently the commutator~\eqref{ssbcomm} would vanish as well. The symmetry breaking vev~$\langle{X_A}\rangle$ is also called order parameter.

We now take the vacuum expectation value of the commutator between the four-current and the scalar field, and we insert the decomposition of the identity as sum over the spectrum of energy-momentum eigenstates~$|n\rangle\equiv\big|\omega\big(\vec{k}_n\big),\vec{k}_n\big\rangle$,\footnote{%
	In our notation the sum over~$n$ is a shorthand for an integration over spatial momenta. Indeed, the orthonormality condition reads $\big\langle\omega\big(\vec{k}_n\big),\vec{k}_n\big|\omega({\scriptstyle\vec{k}_m}),\vec{k}_m\big\rangle=\delta\big(\vec{k}_n-\vec{k}_m\big)$, so that $\sum_n|n\rangle\langle{n}|=\int\!\de\vec{k}_n\;\big|\omega\big(\vec{k}_n\big),\vec{k}_n\big\rangle\big\langle\omega\big(\vec{k}_n\big),\vec{k}_n\big|$.
} that is~$P^\mu_{\vphantom{n}}|n\rangle=k_n^\mu|n\rangle$. At the same time, we translate the fields to the origin, that is
\begin{equation*}
e^{-iP_\mu x^\mu}J^\mu(x)e^{+iP_\mu x^\mu}=J^\mu(0)\ ,
\end{equation*}
and the same for~$X_A$, and we use the translational invariance of the vacuum, $e^{iP\cdot x}|0\rangle=0$, finally obtaining
\begin{align}
&	\Langle\big[J^\mu(x),X_A(y)\big]\Rangle = 	\label{spectrdecompcomm}\\
&\quad =
	\sum_n\Big( e^{ik_n(x-y)}\langle0|J^\mu(0)|n\rangle\langle{n}|X_A(0)|0\rangle -e^{-ik_n(x-y)}\langle0|X_A(0)|n\rangle\langle{n}|J^\mu(0)|0\rangle \Big) \nn
&\quad =
	\int\!\de{k} \sum_n \delta\big(k-k_n\big)\! \Big( e^{ik(x-y)}\langle0|J^\mu(0)|n\rangle\!\langle{n}|X_A(0)|0\rangle -e^{-ik(x-y)}\langle0|X_A(0)|n\rangle\!\langle{n}|J^\mu(0)|0\rangle \!\Big)\ .	\nonumber
\end{align}
The expression in the parenthesis carries a Lorentz vector index, so Lorentz invariance dictates that as a function of momentum it must be proportional to~$k^\mu$. Thus we can define
\begin{align}
\sum_n \delta\big(k-k_n\big)\langle0|J^\mu(0)|n\rangle\langle{n}|X_A(0)|0\rangle &=
	i\left(2\pi\right)^{-d+1} k^\mu \theta\big(k^0\big)\ \rho_{\!A}\big(-\!k^2\big)\ , \label{rhoA}\\
\sum_n \delta\big(k-k_n\big)\langle0|X_A(0)|n\rangle\langle{n}|J^\mu(0)|0\rangle &=
i\left(2\pi\right)^{-d+1} k^\mu \theta\big(k^0\big)\ \tilde{\rho}_A\big(-\!k^2\big)\ , \label{tilrhoA}
\end{align}
where the Heaviside theta step function is there to ensure $k_n$ be the $d$-momentum of a physical state. If both~$J^\mu$ and~$X_A$ are hermitian operators, then the two expressions are just complex conjugates, and $\tilde{\rho}_A\equiv-\rho_A^*$. However, even without assuming this, we can write
\begin{align}
\Langle\big[J^\mu(x),X_A(y)\big]\Rangle &=
	i\left(2\pi\right)^{-d+1}\int\de{k}\ \theta\big(k^0\big)\, k^\mu \left[ e^{ik(x-y)}\rho_{\!A}\big(-\!k^2\big) -e^{-ik(x-y)}\tilde{\rho}_A\big(-\!k^2\big) \right]	\nn
&=
	\left(2\pi\right)^{-d+1} \bfrac{\partial\phantom{x^i}}{\partial{x_\mu}}\int\de{k}\ \theta\big(k^0\big) \left[ e^{ik(x-y)}\rho_{\!A}\big(-\!k^2\big) +e^{-ik(x-y)}\tilde{\rho}_A\big(-\!k^2\big) \right]	\nn
&=	
	\bfrac{\partial\phantom{x^i}}{\partial{x_\mu}}\int\de{m^2} \left[ \rho_{\!A}\big(m^2\big)\Delta^{\!+\!}(x-y) +\tilde{\rho}_A\big(m^2\big)\Delta^{\!+\!}(y-x) \right]\ ,	\label{currscalcomm}
\end{align}
where we have introduced the standard positive frequency scalar distribution,
\begin{equation}	\label{freescalcorr}
\Delta^{\!+\!}(x-y)= 	\left(2\pi\right)^{-d+1} \int\de{k}\ \theta\big(k^0\big)\, \delta\big(k^2+m^2\big)\; e^{ik(x-y)}\ .
\end{equation}
We remind the reader that for space-like separations we have $\Delta^{\!+\!}(x-y)\equiv\Delta^{\!+\!}(y-x)$, so that the canonical commutator for a complex scalar field~$\phi$,
\begin{equation} 	\label{scalcomm}
\Langle\big[\phi(x),\phi^\dagger(y)\big]\Rangle=\Delta(x-y)\equiv \Delta^{\!+\!}(x-y)-\Delta^{\!+\!}(y-x)
\end{equation}
vanishes at space-like separations, respecting causality. In the same way, the commutator~\eqref{currscalcomm} has to vanish at space-like separations. This implies $\rho_{\!A}\big(m^2\big)=-\tilde{\rho}_A\big(m^2\big)$, so giving
\begin{equation}	\label{commDelta}
\Langle\big[J^\mu(x),X_A(y)\big]\Rangle =	
	\bfrac{\partial\phantom{x^i}}{\partial{x_\mu}}\int\de{m^2}\ \rho_{\!A}\big(m^2\big)\; \Delta(x-y)\ .
\end{equation}
Now we can use the current conservation, and, recalling $\Delta(x-y)$ is a solution of the Klein-Gordon equation, we can write
\begin{equation*}
0= \bfrac{\partial\phantom{x^i}}{\partial{x^\mu}}\Langle\big[J^\mu(x),X_A(y)\big]\Rangle =
	\Box_x\int\de{m^2}\ \rho_{\!A}\big(m^2\big)\; \Delta(x-y) =
		\int\de{m^2}\ m^2\, \rho_{\!A}\big(m^2\big)\; \Delta(x-y)\ .
\end{equation*}
For time-like or light-like separations, this yields the condition
\begin{equation}	\label{rhoisdelta}
m^2\rho_{\!A}\big(m^2\big) = 0\ ,
\end{equation}
which implies that $\rho_{\!A}$ must vanish for any value of~$m^2$, except~$m^2=0$. From the definition of~$\rho_{\!A}$~\eqref{rhoA}, it is clear that this latter possibility occurs only if there is a state~$|\bar{n}\rangle$ in the spectrum that has zero mass, \emph{i.e.}:~$-k_{\bar{n}}^2=0$. 

We now have to see that in case of spontaneously broken symmetry the function~$\rho_{\!A}$ actually cannot vanish everywhere, implying the existence of this massless mode. Let us take the expression~\eqref{commDelta} for the temporal component, and let us integrate it and evaluate it at equal times~$x^0=y^0=t$, in order to obtain the symmetry breaking condition~\eqref{ssbcomm}. We find
\begin{align*}
\!\!
-iT_{AB}\Langle X_B\Rangle &=
	\int\!\de{m^2}\rho_{\!A}\big(m^2\big)\int\!\de\vec{x} \int\!\bfrac{\de{k}\,(ik^0)}{(2\pi)^{d-1}}\,\theta\big(k^0\big)\delta\big(k^2+m^2\big)\Big(e^{i\vec{k}\cdot\vec{x}}+e^{-i\vec{k}\cdot\vec{x}}\Big) \\
&=	
	i\int\de{m^2}\ \rho_{\!A}\big(m^2\big) \int\de\vec{x}\int\!\!\bfrac{\!\de\vec{k}\;\;\;}{(2\pi)^{d-1}}\ e^{i\vec{k}\cdot\vec{x}} =
		i\int\de{m^2}\ \rho_{\!A}\big(m^2\big)\ ,
\end{align*}
where we have used the distributional identity
\begin{equation}	\label{deltaident}
\theta\big(k^0\big)\, \delta(k^2+m^2) = \bfrac{1}{2k^0}\; \delta\Big(k^0-\sqrt{\smash[t]{\vec{k}^{\,2}+m^2}\vphantom{\big|}}\,\Big)\ .
\end{equation}
We then conclude
\begin{equation*}
\rho_{\!A}\big(m^2\big)= -T_{AB}\Langle X_B(0)\Rangle\ \delta(m^2)\ ,
\end{equation*}
as announced, implying the existence of a massless state in the spectrum. Now, a delta function contribution can only be generated by a single-particle pole. Furthermore, in order to contribute to~$\rho_{\!A}$, such single-particle state has to couple both to the conserved current and to the scalar operator that gets a vev, otherwise either $\langle0|J^\mu(0)|n\rangle$ or $\langle{n}|X_A(0)|0\rangle$ would respectively vanish. The latter condition implies that our single-particle state must be a scalar (spin~$0$) particle; the former entails that it inherits the parity and internal quantum numbers of~$J^0$. This precise gentleman is the notorious Nambu-Goldstone boson.

The proof outlined here has been carried out in generic $d+1$ space-time dimensions. However, by an argument first raised by Coleman~\cite{Coleman:1973ci}, there is an obstruction to spontaneous symmetry breaking in $(1+1)$-dimensions. Actually, nothing really breaks down in the steps of the proof of Goldstone theorem, it is rather the possibility for a scalar operator to get a stable vacuum expectation value that is ruled out in two dimensions, so invalidating the very initial point of the reasoning~\eqref{ssbcomm}. In next section we present Coleman's mathematical argument and discuss its physical implications.

\subsubsection[Coleman theorem]{Lowering to $\boldsymbol{1\!+\!1}$ dimensions: Coleman theorem}
\label{ColemanTh}

The key feature, on which Coleman's argument is founded, is the fact that in one spatial dimension the scalar two-point correlator~\eqref{freescalcorr} has a logarithmic behavior in position space, so having an infrared divergence for massless scalar fields:
\begin{align}
\Langle\phi(x)\phi^\dagger(y)\Rangle &= \Delta^{\!+\!}_{d=1}(x-y) = 
	\left(2\pi\right)^{-1} \int\de^2{p}\ \theta\big(k^0\big)\, \delta\big(k^2+\mu^2\big)\; e^{ik(x-y)} = \nn
& =	
	\bfrac{1}{2\pi} \int\frac{\de{k^1}\ e^{ik(x-y)}}{2\sqrt{\!{(k^1)}^2+\mu^2}} = 
		-\bfrac{1}{4\pi}\ \ln\big(-\mu^2(x-y)\big)\ ,	\label{logDelta}
\end{align}
where $\mu^2$ is indeed an infrared regulator to escape the divergence of zero mass.

Rather than reproduce Coleman's proof~\cite{Coleman:1973ci}, which is quite mathematical, we will follow the lines of Ma and Ranjaraman~\cite{Ma:1974tp}, which present a more physical picture of what happens in a two dimensional quantum field theory with an internal continuous symmetry. Moreover, this picture will allow to better understand some situations that seem to evade Coleman's argument, in particular the strict large~$N$ limit of~SU$(N)$ field theories, such as the Thirring model~\cite{Thirring:1958in}, or the holographic dual theory of reference~\cite{Argurio:2016xih}.

We consider a U$(1)$-invariant Lagrangian for a complex scalar field, with an arbitrary (U$(1)$-invariant, and bounded from below) potential, and we write the field in the polar representation, $\phi=\frac{1}{\sqrt{2}}\,\rho\,e^{i\vth}$:
\begin{equation}
\CL= -\partial^\mu\phi^*\partial_\mu\phi -V(\phi^*\phi) = 
	-\tfrac12\partial^\mu\rho\partial_\mu\rho -\tfrac12\rho^2\partial^\mu\vth\partial_\mu\vth -V\big(\rho^2\big)\ .
\end{equation}
This Lagrangian can be regarded of course as a low-energy effective one of a more general and involved theory. The vacuum expectation value for this scalar field is dictated by the minimum of the potential in phase space. If we want to have spontaneous symmetry breaking, we need the minimum to be away from the origin~$\rho=0$. We take then $\rho^2=|v|^2$, with $v\in\mathbb{C}$. Since the potential is invariant under the global~U$(1)$, our minimum in~$|v|^2$ represents a continuity of possible vacua for~$\phi$, varying the phase of~$v$. In higher dimensions, the selection of a specific vacuum, that is of a specific phase, yields the spontaneous breaking. In $(1+1)$-dimensions, the IR divergences of the scalar propagator bring large quantum fluctuations of the phase field~$\vth$, that would spread away any classically selected phase. Thus, the classical possibility of picking a vacuum out of the continuum is ruled out at the quantum level.

Let us see this fact explicitly. By making $\rho=v$ non-dynamical, we are left with the following Lagrangian for $\vth$,
\begin{equation}	\label{vthLagr}
\CL= -\tfrac12v^2\partial^\mu\vth\partial_\mu\vth\ .
\end{equation}
We then consider the expansion of the scalar field~$\vth$ in its annihilation and creation components, $\vth=\vth^+ + \vth^-$, with
\begin{equation}
\vth^+(t,x)=\int\bfrac{\:\de{k}\ e^{-i\omega t+ikx}}{v\,\sqrt{4\pi\omega}}\ a_k^{\phantom{\dagger}}\ , \quad
	\vth^-(t,x)=\int\bfrac{\:\de{k}\ e^{i\omega t-ikx}}{v\,\sqrt{4\pi\omega}}\ a_k^\dagger\ ,
\end{equation}
where $(k^0,k^1)$ have been replaced by~$(\omega,k)$, and we have an extra $v$-factor because of the form of the kinetic term in the Lagrangian~\eqref{vthLagr}. Then, using the standard commutations relations for annihilation/creation operators, we straightforwardly have
\begin{equation}	\label{vthcomm}
\Langle\big[\vth^+(x),\vth^-(0)\big]\Rangle =
	\int\bfrac{\:\de{k}\ e^{ikx}}{4\pi v^2\sqrt{k^2}} = 
		\bfrac{1}{v^2}\lim_{\mu^2\to0}\Delta^{\!+\!}_{d=1}(x)\ ,
\end{equation}
which, from~\eqref{logDelta}, is diverging to positive infinity. So, we can evaluate
\begin{equation*}
\!\Langle e^{i\vth(x)}\Rangle =
	\Langle e^{i\left(\vth^-(x)+\vth^+(x)\right)}\Rangle = 
		\Langle e^{i\vth^-(x)}e^{i\vth^+(x)}e^{-\frac12\left[\vth^+(x),\vth^-(x)\right]}\Rangle =
			e^{-\frac{1}{2v^2}\Delta^{\!+\!}_{d=1}(0)} =0 \ ,
\end{equation*}
where we have used the Baker–Campbell–Hausdorff formula for exponentials of operators, and the annihilation property of the vacuum. Thus, we have shown that the expectation value of~$e^{i\vth(x)}$ is wiped out by the diverging quantum fluctuations, and so $\Langle\cos{\vth(x)}\Rangle$ and~$\Langle\sin{\vth(x)}\Rangle$ vanish as well. No preferred direction in the phase plane survives the quantum smearing in two dimensions.

However, if we had an arbitrary small, but non zero explicit breaking term (such for instance a small mass for~$\vth$), it would act precisely as the regulator~$\mu^2$ in~\eqref{logDelta} and make the commutator~\eqref{vthcomm} finite. The concomitant occurrence of spontaneous and explicit breaking, with explicitly breaking parameter hierarchically suppressed with respect to the spontaneously breaking vev, gives rise to so-called pseudo-Goldstone bosons. Pseudo-Goldstone bosons are would-be Goldstone particles that have a mass which is hierarchically lower with respect to the rest of the spectrum, thanks to the mentioned hierarchy between explicit and spontaneous breaking. 

This last scenario will be discussed in depth in next section, but, before moving to that, let us discuss a class of model that seems to evade Coleman theorem, that is SU$(N)$-invariant theories in the large~$N$ limit. Of course we are interested in large~$N$ theories in view of holographic realizations of symmetry breaking, where pure boundary computations correspond to the strict infinite~$N$ limit.

So now we consider the SU$(N)$ extension of Thirring model~\cite{Thirring:1958in}, proposed by Gross and Neveu~\cite{Gross:1974jv}, with our focus on understanding the peculiar features of the infinite~$N$ limit for spontaneous symmetry breaking in two dimensions, rather than on the details of the model. This is a theory of $N$-component massless fermions with quartic interaction in two dimensions, which has a smooth, weak-interacting limit for~$N\to\infty$, so that it can be solved in a $1/N$~expansion. On top of the SU$(N)$ invariance, the system enjoys a U$(1)$ chiral symmetry, which normally would prevent the fermions from dynamically getting a mass. However, as pointed out by Gross and Neveu, the theory allows for a minimum of the potential away from the origin in the phase plane, fermions fields acquire mass, and a massless boson arises: so, everything suggests that spontaneous symmetry breaking is occurring in two dimensions.

Nonetheless, Witten showed~\cite{Witten:1978qu} that the physical fermions are actually not charged under the chiral symmetry, explaining why they are not protected from having a mass, and that no long-range order is present, so no symmetry breaking and no contradiction to Coleman theorem. Indeed, the low-energy effective Lagrangian for Gross-Neveu-Thirring model is of the form~\cite{Witten:1978qu}
\begin{equation}
\CL= -\frac{N}{4\pi}\,\partial^\mu\vth\partial_\mu\vth\ ,
\end{equation}
quite analogous to~\eqref{vthLagr}. Then we have
\begin{equation}
\Langle e^{-i\vth(x)}e^{+i\vth(0)}\Rangle =
	e^{\frac{\pi}{N}\left(\Delta^{\!+\!}_{d=1}(x)-\Delta^{\!+\!}_{d=1}(0)\right)} \sim x^{-\frac{1}{N}} \ ,
\end{equation}
for large~$x$. So the long-range order is suppressed for~$x\to\infty$. In a spontaneously broken theory, this expectation value would be finite (proportional to the vev), even for infinite separations, as in~\eqref{ssbcomm}. For unbroken symmetry, instead, we would have an exponential fall-off, at the rate of the lightest excitation in the spectrum: $\Langle e^{-i\vth(x)}e^{+i\vth(0)}\Rangle\sim e^{-mx^2}$. So, this two-dimensional large~$N$ case is in-between: the long range order vanishes, but with a power law, so less quickly than in an usual unbroken phase.

This was the main result of Witten's paper~\cite{Witten:1978qu}, but what he did not stress in that work, since it was not yet relevant at that time, is the fact that in the strict infinite~$N$ limit the long-range order is restored. And indeed, in the holographic setup of Section~\ref{HoloColeman}, from a pure boundary analysis (which is at strictly infinite~$N$), we retrieve the Ward identities of spontaneous symmetry breaking, as in any higher dimensions.

In next section we derive in full generality such Ward identities, which are a non-per\-tur\-ba\-tive feature of symmetry breaking in quantum field theory. Furthermore, as already announced, we will consider the concomitant occurrence of spontaneous and explicit breaking, and analyze the physical predictions that can be extracted just from Ward identities, independently of any specific theory.

\subsection{Explicit breaking and Ward identities}
\label{explicitWI}

We consider here an action that is invariant under an internal global continuous symmetry, as in~\eqref{invS}, and we add a term that breaks the symmetry explicitly:
\begin{equation}
S_\mathrm{tot} = S_\mathrm{inv} + S_m\ , \quad
\text{with }\ S_m = \bfrac12\, m\intd{} \big(\CO[X] + \CO^{*\!}[X]\big) = m\intd{} \ReO[X]\ ,
\end{equation}
where $\CO_\phi$ is a scalar operator of scaling dimension $\Delta$,  which is charged under the symmetry, and does not depend on derivatives of the fields. For the sake of simplicity, and for an easier comparison with the results of the holographic model of Section~\ref{HoloPGB}, we choose as global symmetry an abelian U$(1)$, and we assume
\begin{equation}
\vardel_\alpha\CO= i\alpha\:\CO\ , \quad\Rightarrow\ \left\{\begin{array}{ll}
	\vardel_\alpha\ReO=-\alpha\ImO 	\!\!\!\!\!& , \\
	\vardel_\alpha\ImO=+\alpha\ReO 	\!\!\!\!\!& .
\end{array} \right.
\end{equation}

Since $\vardel_\alpha S_{\mathrm{inv}}=0$, as we have seen in Section~\ref{GoldstoneSSB}, for $m=0$ we have a conserved current~$J^\mu$. But in presence of the explicit breaking term, the equations of motion get modified accordingly,
\begin{equation*}
\partial_\mu\bfrac{\delta\CL_\mathrm{inv}}{\delta\partial_\mu{X_A}} -\bfrac{\delta\CL_\mathrm{inv}}{\delta{X_A\;\;}}  -m\,\bfrac{\delta\ReO}{\delta{X_A\;\;}} =0\ , 
\end{equation*}
so that in the second line of~\eqref{divJvanish} we now have
\begin{align}
0=\vardel_\alpha S_\mathrm{inv} &=  
	\intd{} \bigg[ -m\,\bfrac{\delta\ReO}{\delta{X_A\;\;}}\,\vardel_\alpha X_A +\partial_\mu\bigg(\bfrac{\delta\,\CL\quad}{\delta\partial_\mu{X_A}}\,\vardel_\alpha X_A\bigg) \bigg] \nn
&=	
		-\intd{}\, \alpha\, \big( \partial_\mu J^\mu -m\,\ImO \big)	\label{nonconscurr}
\end{align}
where we have used the trivial identity~$\bfrac{\delta\CO\;}{\delta{X_A}}\vardel_\alpha X_A\equiv\vardel_\alpha\CO$. We see that the current is no longer conserved, instead we have the operator identity
\begin{equation}	\label{deJmImO}
\partial_{\mu}J^\mu=m\, \ImO\ .
\end{equation}

We now want to show that such identity holds at the quantum level, and derive the Ward identities corresponding to the breaking of an internal symmetry. For that we employ the path integral formulation, and we use the invariance of the measure under the symmetry transformation.

The path integral and $n$-point time-ordered correlation functions are defined as follows:
\begin{equation}	\label{pathcorr}
\mathcal{Z} = \int\!\mathcal{D}[X]\: e^{iS_\mathrm{tot}[X]} \ , \qquad
	\Langle\mathcal{F}[X](x_1,\dots,x_n)\Rangle = \int\!\mathcal{D}[X]\: e^{iS_\mathrm{tot}[X]}\, \mathcal{F}[X](x_1,\ldots,x_n)\ .
\end{equation}
If there are no quantum anomalies (violation of the symmetry at the quantum level), the invariance of the measure~$\mathcal{D}[X]$ under field redefinitions assures
\begin{equation}	\label{measureinv}
\int\!\mathcal{D}[X]\ \vardel_\alpha\Big(e^{iS_\mathrm{tot}[X]}\:\mathcal{F}\Big) =0\ , \qquad \forall\ \mathcal{F}\ .
\end{equation}
This is then true for the path integral itself, allowing to confirm the operator identity~\eqref{deJmImO} at the quantum level. Indeed, from~\eqref{measureinv} we have
\begin{equation}	\label{varStot0}
0= \int\!\mathcal{D}[X]\ \vardel_\alpha e^{iS_\mathrm{tot}} = \int\!\mathcal{D}[X]\ e^{iS_\mathrm{tot}}\;i\,\vardel_\alpha S_\mathrm{tot}\ .
\end{equation}

In addition, in order to compute correlation functions involving the divergence of the current, which classically vanishes, we need to make the global symmetry local. In this way the variation of the invariant action is not zero anymore, rather $\vardel_\alpha S_\mathrm{inv}=-\intd{}\,J^\mu\partial_\mu\alpha$, and consequently
\begin{equation}	\label{totlocvar}
\vardel_\alpha S_\mathrm{tot}=\vardel_\alpha S_\mathrm{inv}+\vardel_\alpha S_m =
-\intd{}\,\left(J^\mu\partial_\mu\alpha+\alpha\,m\ImO\right)\ ,
\end{equation}
A perfectly equivalent approach is to add to the path integral sources for both the current and~$\ImO$, whose variations compensate the variation of the total action~\eqref{totlocvar}. The source for the current then has to be a vector field~$A_\mu$ transforming under U$(1)$ gauge transformations, $\vardel_\alpha A_\mu=\partial_\mu\alpha$; the source for~$\ImO$ has to be a scalar field~$m$ transforming under the~U$(1)$ as $\vardel_\alpha m=m\alpha$. Thus we have
\begin{equation}	\label{pathintsources}
\int\!\mathcal{D}[X]\ e^{iS_\mathrm{inv}+iS_\mathrm{m}+i\!\int\!A_{\mu}J^\mu +i\!\int\!m\ImO}\ .
\end{equation}

Now we can choose either to put the total variation~\eqref{totlocvar} into the vanishing expression~\eqref{varStot0}, or to vary the sourced path integral~\eqref{pathintsources} with respect to the sources only, and require it to be invariant (since transformations of the sources can be absorbed by transformations of the fields),
\begin{align*}
0 &= 
	\int\!\mathcal{D}[X]\ e^{iS_\mathrm{inv}+iS_\mathrm{m}+i\!\int\!A_{\mu}J^\mu +i\!\int\!m\ImO} i\!\intd{d}\Big(J^\mu\vardel_\alpha A_{\mu}+\ImO\vardel_\alpha m\Big) = \\
&=
	\intd{d}\Langle J^\mu\partial_\mu\alpha+\alpha\,m\ImO \Rangle\ .
\end{align*}
In both ways, after integration by parts, we obtain the quantum version of the operator identity~\eqref{deJmImO}:
\begin{equation}
\Langle\partial_{\mu}J^\mu\Rangle=m\,\Langle\ImO\Rangle\ .
\end{equation} 

Analogously, we can start with the variation of the path integral for the operator~$\ImO$,
\begin{align*}
0 &= 
	\int\!\mathcal{D}[X]\ \vardel_\alpha\Big(e^{iS_\mathrm{tot}}\ImO(x\p)\Big) = 
		\int\!\mathcal{D}[X]\ e^{iS_\mathrm{tot}}\left( i\,(\vardel_\alpha S_\mathrm{tot})\ImO(x\p)+\vardel_\alpha\ImO(x\p) \right) = \\
&=
	\int\!\mathcal{D}[X]\ e^{iS_\mathrm{tot}} \left[ \alpha(x\p)\ReO(x\p) +i\intd{}\alpha(x)\left(\partial_{\mu}J^\mu(x)-m\ImO(x)\right)\ImO(x\p) \right]\ ,
		\label{pathintvarImO}
\end{align*}
thus obtaining the following Ward identity for symmetry breaking:
\begin{equation}	\label{WIsb}
\Langle\partial_{\mu}J^\mu(x)\ImO(0)\Rangle = m\,\Langle\ImO(x)\ImO(0)\Rangle +i\,\Langle\ReO\Rangle\, \delta(x)\ ,
\end{equation}
We see that, if spontaneous breaking occurs as well, by the real part of the scalar operator taking a non-zero vacuum expectation value, $\Langle\ReO\Rangle=\!v$, we have a contact term, which survives in the $m\to0$~limit, 
\begin{equation}	\label{WIssb}
\Langle\partial_\mu J^\mu(x)\ImO(0)\Rangle = iv\;\delta(x)\ ,
\end{equation}
that is even when the current is perfectly conserved at the classical level. Then, Lorentz invariance imposes, Fourier transforming to momentum space,
\begin{equation}	\label{GBpoleWI}
\Langle J^\mu(k)\ImO(-k)\Rangle = iv\ \frac{k^\mu}{k^2}\ ,
\end{equation}
manifesting the presence of a massless pole in the spectrum, signature of the Goldstone boson. Requiring continuity for $m$ going to zero in the Ward identity~\eqref{WIsb}, the same massless pole has to appear in the two-point scalar correlator~$\Langle\ImO\ImO\Rangle$ as well, although its explicit computation requires an analysis of the (IR) dynamics, and so knowledge over the specific theory. 

When $m\neq0$, instead, thanks to the Ward identity~\eqref{WIsb} and to the operator identity~\eqref{deJmImO}, we have that the following two-point functions all depend on a single non-trivial function $f(\Box)$:
\begin{align}
\Langle\ImO \ImO\Rangle & = -i\,f(\Box)\ ,	\label{oocorr}\\
\Langle\partial_\mu J^\mu \ImO\Rangle & = -im\, f(\Box) +iv\ ,	\label{jocorr}\\
\Langle\partial_\mu J^\mu \partial_\nu J^\nu\Rangle & = -im^2 f(\Box) +imv\ ,	\label{jjcorr}
\end{align}
where we have kept the delta function implicit. The last correlator is just a consequence of the operator identity, and when $v=0$ (pure explicit breaking case) then also the second correlator is a trivial consequence of the operator identity. 

When both $v\neq 0$ and $m\neq 0$, we see that the scalar-current correlator~\eqref{jocorr} exhibits both features: a term related to \eqref{oocorr} and a constant term. On the other hand, since the symmetry is broken explicitly, we do not expect a massless mode in the spectrum contributing to this set of correlators. However, in the case where $m\ll v^{\bfrac{d-\Delta}{\Delta\vphantom{|}}}$ (small amount of explicit breaking compared to spontaneous breaking), we expect to find a light particle, whose mass is sensibly lower with respect to the rest of the spectrum and goes to zero in the $m\to0$ limit. We call such particle pseudo-Goldstone boson, precisely in consequence of this fact that for~$m\to0$ it `becomes' the Goldstone boson. Furthermore, the leading-order contribution to the square mass of the pseudo-Goldstone for small explicit breaking is linear in~$\frac{m}{v}$, as we will show now in full generality, requiring continuity in the $m\to0$ limit. This linear relation for the square mass of the pseudo-Goldstone boson is renown as GMOR (Gell-Mann--Oakes--Renner) relation~\cite{GellMann:1968rz}.

The GMOR relation was first found in the context of the effective description of pions as pseudo-Goldstone bosons of the approximate chiral symmetry of quantum chromodynamics, where the role of explicit breaking parameter is played by the masses of the quarks. Standard derivations of GMOR for the pion can be found in the literature, for instance in~\cite{Gasser:1982ap,Giusti:1998wy}, and also~\cite{Evans:2004ia} for a derivation based on the effective action. Here we derive it in a more general fashion, as it was done in~\cite{Argurio:2015wgr}.

In momentum space we can write
\begin{equation}	\label{WImomsp}
\Langle \ImO \ImO\Rangle = -i\,f\big(k^2\big)\ , \qquad 
	ik_\mu \Langle J^\mu\ImO\Rangle = -im\,f\big(k^2\big) +iv\ .
\end{equation}
Note that  the dimensions are  $[f]=2\Delta-d$, $[v]=\Delta$ and $[m]=d-\Delta$. Using Lorentz invariance, which imposes~$\Langle{J}^\mu\ImO\Rangle=k^{\mu}g\big(k^2\big)$, the second relation in~\eqref{WImomsp} leads to
\begin{equation}	\label{JmuImOnospurious}
\Langle  J^\mu \ImO\Rangle = -\bfrac{k^\mu}{k^2}\:\Big(m\,f\big(k^2\big)-v\Big)\ .
\end{equation}
We immediately see that there cannot be a massless excitation in the $\Langle \ImO \ImO\Rangle$ channel, otherwise there would be a double pole in $\Langle  J^\mu \ImO\Rangle$. Moreover, the massless pole in the above correlator should be spurious, which requires $f\big(k^2\big)$ to satisfy
\begin{equation}	\label{scheme}
m\,f(0)-v=0\ .
\end{equation}

So far, everything is valid for any values of~$m$ and~$v$. When $m^2$ and~$k^2$ are both small with respect to~$v$, we can approximate~$f$ by a pole of mass~$M$, corresponding to the pseudo-Goldstone mode,
\begin{equation}
f\big(k^2\big)\ \simeq \ \frac{\mu}{k^2+M^2} -\frac{\mu}{M^2} +\frac{v}{m}\ ,\label{fpgb}
\end{equation}
where we have implemented the condition \eqref{scheme}, and the residue $\mu$ is a dynamical quantity of dimension $2\Delta-d+2$. 

We now require that in the $m\to 0$ limit $f\big(k^2\big)$ go over smoothly to what we expect in the pure spontaneously broken case. Namely, we expect $\mu$ to be (roughly) constant in the limit, as of course $v$, while $M^2 \to 0$, so that 
\begin{equation}
f\big(k^2\big) \;\xrightarrow{m\to0}\; \frac{\mu}{k^2}\ ,
\end{equation}
up to possibly an additive finite constant. From \eqref{fpgb} we see that this is possible only if there is a relation between all the constants such that 
\begin{equation}
M^2 = \bfrac{\mu}{v}\; m \ .\label{gmor}
\end{equation}

This is the generalized form of the GMOR relation \cite{GellMann:1968rz}, which indeed states that, at first order, the square mass of the pseudo-Goldstone boson scales linearly with the small parameter which breaks explicitly the symmetry, as announced. The two other constants entering the expression are both of the order 
of the dynamical scale generating the vev, \emph{i.e.} the spontaneous breaking of the symmetry. \label{vrealmreal}We remark that, since $\mu$ has to be positive because of unitarity, then the signs (and more generally the phases) of $m$ and $v$ have to be correlated in order to avoid tachyonic pseudo-Goldstone bosons. This can be understood by the fact that the small explicit breaking removes the degeneracy of the vacua, and thus the phase of the vev $v$ is no longer arbitrary but has to be aligned with the true vacuum selected by $m$. 

We conclude making the link with the usual form in which GMOR relation is stated, which is in terms of the residue of the full current-current correlator,
\begin{equation}	\label{currcurrcorr}
\Langle J^\mu J^\nu\Rangle =- \frac{i\mu_J}{k^2+M^2}k^\mu k^\nu +\dots\ ,
\end{equation}
where the residue~$\mu_J$ is actually given by the square of the decay constant of the pion,~$f_\pi^2$. Then we can use the GMOR value of the pole~\eqref{gmor} into~\eqref{fpgb}, and rewrite the longitudinal correlator~\eqref{jjcorr} as
\begin{equation}	\label{longcurrcorr}
k_\mu k_\nu \Langle J^\mu J^\nu\Rangle = \frac{imv\,k^2}{k^2+\mu\frac{m}{v}}\ .
\end{equation}
The comparison of these two expressions,~\eqref{currcurrcorr} and~\eqref{longcurrcorr}, at the location of the pseudo-GB pole~\eqref{gmor} yields
\begin{equation}
\mu_J = \frac{v^2}{\mu}\ ,
\end{equation}
and so
\begin{equation*}
M^2 = \bfrac{v}{\mu_J}\; m \equiv \bfrac{v}{f_\pi^2}\; m \ ,
\end{equation*}
which is the desired usual form of GMOR relation~\cite{GellMann:1968rz}, though completely equivalent to \eqref{gmor}.

Here we have kept both $d$ and $\Delta$ arbitrary, and the relation is valid in all generality. In Section~\ref{HoloPGB} we will retrieve this relation in a particular holographic model for $d=3$ and $\Delta=2$, determining as well the value of the residue~$\mu$, which is specific to the considered model.

\subsection[Goldstone theorem in non-relativistic field theories]{Goldstone theorem in non-relativistic field theories}	\label{nonrelGoldstoneTh}

In our derivation of Goldstone theorem, we have used Lorentz invariance in defining the spectral density~(\ref{rhoA}-\ref{tilrhoA}), as well as in any occurrence of the Lorentz-invariant distributions~$\Delta^+(x)$ and~$\Delta(x)$. So, in lack of Lorentz invariance, that proof does not apply. However, the central statement, that spontaneously broken symmetries imply zero-energy modes in the spectrum, holds, in a softer formulation and with some modifications, also for non-relativistic field theories, as we will see.

On the other hand, though in Section~\ref{GoldstoneSSB} we have presented Noether theorem as well in a relativistic formalism, it relies only on the action principle and so it holds independently of the symmetries of space-time. However, the key assumption in order to have a conserved charge is that the spatial divergence of the current vanishes at spatial infinity. Actually, to ensure time-independence of the vev in the spontaneous breaking condition~\eqref{ssbcomm}, it is sufficient that
\begin{equation}	\label{spatdivcurr}
\int\!\de{\vec{x}}\ \big[\partial_i^{\vphantom{a}}J^a_i(\vec{x}),X_A\big] = 0\ ,
\end{equation}
so that $\partial_t\big[Q^a,X_A\big]=0$. In relativistic theories, which are intrinsically causal, this is guaranteed as long as the operator~$X_A$ is localized to a finite domain of space-time, and Goldstone theorem inescapably follows\footnote{%
	Goldstone theorem predicts a massless particle, but this does not mean that such particle be physically observable. As high-energy physicists we all have in mind the example of gauge theories and Brout-Englert-Higgs mechanism, where the Goldstone boson does not appear in the observable spectrum, being `hidden' in the longitudinal degree of freedom of the newly massive gauge boson.
}. On the contrary in non-relativistic theories, which can be non-causal, long-range interactions can in some cases make the boundary term in~\eqref{spatdivcurr} not vanish. It can be shown that~\eqref{spatdivcurr} holds as long as the range of interactions is finite~\cite{Lange:1965zz,Lange:1966zz}, otherwise the applicability of condition~\eqref{spatdivcurr} has to be checked case by case. In the following, we assume~\eqref{spatdivcurr} as necessary prerequisite for spontaneous breaking to occur, otherwise the charge commutator and vev itself are not time-independent, and Goldstone theorem cannot be discussed at all.

We can then take the symmetry breaking commutator~\eqref{ssbcomm}, and rewrite it through the spectral decomposition of the unity as in~\eqref{spectrdecompcomm}, obtaining
\begin{align}
&\Langle\big[Q^a,X_A(0)\big]\Rangle =	\label{chargecomm}\\
&\qquad\ \displaystyle	=
	\sum_n\int\de\vec{x}\ \Big( e^{ik_nx} \Langle{J^a_t(0)}\big|n\big\rangle\Langle{n}\big|X_A(0)\Rangle -e^{-ik_nx}\Langle{X_A(0)}\big|n\big\rangle\Langle{n}\big|J^a_t(0)\Rangle \Big)\ \phantom{.} \nn
&\qquad\ \displaystyle	=
	(2\pi)^{d}\sum_n\delta(\vec{k}_n) \Big( e^{-i\omega(\vec{k}_n)t} \Langle{J^a_t(0)}\big|n\big\rangle\Langle{n}\big|X_A(0)\Rangle -e^{i\omega(\vec{k}_n)t}\Langle{X_A(0)}\big|n\big\rangle\Langle{n}\big|J^a_t(0)\Rangle \Big)\ . \vphantom{\int^l}	\nonumber
\end{align}
We see that, for the left-hand side to be time-independent, the right-hand side must vanish except for states with vanishing energy at zero momentum, that is: $\omega\big(\vec{k}_{n\!}\!\equiv\!0\big)=0$~\cite{Guralnik:1967zz}. Furthermore, since the left-hand side is non-zero because of the spontaneous breaking condition~\eqref{ssbcomm}, \emph{at least one} of such states must exist and contribute to the right-hand side. So, even without using any Lorentz invariance, we can state that, when an internal continuous symmetry is spontaneously broken, there exists at least one mode, whose energy goes to zero for spatial momentum going to zero, and which couples to both the conserved current and the scalar operator that does not commute with the current.

Let us now point out the crucial differences between the general statement of Goldstone theorem, which we have just outlined here, and the relativistic one. First, in its more general version Goldstone theorem is a low-energy statement: for $\vec{k}\to0$ there is a vanishing energy mode, but it is not assured to survive at high energy, and in any case not with the same properties. In the relativistic case, the dispersion relation (\emph{i.e.}~the dependence of the energy of an excitation on the momentum) is fixed to~$\omega_{\vec{k}}=\mod$, at low as at high energy, and the Goldstone mode is an actual massless particle.

In a non-relativistic framework a richer range of dispersion relations are recorded. Let us assume broken Lorentz invariance, but still preserved invariance under spatial rotations. Then, in such case the Fourier transform of the old same correlator~\eqref{spectrdecompcomm} has a less constrained structure than in the Lorentz invariant case, which we can express in a formally Lorentz covariant way using a time-like vector~$b^\mu=(1,\vec{0})$:
\begin{equation}	\label{nonrelcomm}
\intd{ } e^{ikx}\Langle\big[J^\mu(x),X_A(0)\big]\Rangle =
	k^\mu\rho_1\big(k^2,b\cdot{k}\big) + b^\mu\rho_2\big(k^2,b\cdot{k}\big)\ .
\end{equation}
In the relativistic case, only the term proportional to~$k^\mu$ was present~(\ref{rhoA}-\ref{tilrhoA}). Then the current conservation leads to the non-relativistic equivalent~\cite{Gilbert:1964iy,Guralnik:1967zz} of condition~\eqref{rhoisdelta}, which constrains~\eqref{nonrelcomm} to take the form
\begin{equation}
k^\mu\:\delta\big(k^2\big)\,\tilde{\rho}_1\big(b\cdot{k}\big)
	+\left(k^2b^\mu-(b\cdot{k})k^\mu\right)\tilde{\rho}_2\big(k^2,b\cdot{k}\big)
		+b^\mu\:\delta\big(b\cdot{k}\big)\,\tilde{\rho}_3\big(k^2\big)
			+b^\mu\;C_4\delta\big(k^0\big)\delta\big(\vec{k}\big)\ ,	\label{nonrelstruct}
\end{equation}
where $C_4$ is just a constant, and the $\tilde{\rho}$'s are functions only of the explicitly indicated arguments. We can see that the introduction of the second term in~\eqref{nonrelcomm} brings quite more possible contributions, which in the early stages had been advocated as signals of a breakdown of Goldstone theorem in the non relativistic framework~\cite{Gilbert:1964iy,Klein:1964ix}. 

However, the last two terms correspond to states with zero energy, and zero energy and zero spatial momentum, respectively. The latter is a spurious isolated energy-momentum eigenstates with eigenvalue~$k^\mu=0$, whose contribution can be excluded if forces have finite (exponentially suppressed) range~\cite{Lange:1966zz,Nielsen:1975hm}~(see also the discussion around eq.~(32) of Brauner's review~\cite{Brauner:2010wm}). 

The former can be one of the degenerate ground states of the continuum of possible vacua, which are not actual excitations of the spectrum with energy going to zero \emph{in the limit~$\mod\to0$}. The possibility of such term contributing to~\eqref{nonrelcomm} is an artifact of working at infinite volume, and it is ruled out by restricting to a finite volume. In a finite volume, indeed, the integration in~\eqref{chargecomm} does not give exactly a delta, rather a function peaked around zero, but which allows for values of~$\mod$ slightly different from zero. So, at finite volume, the states contributing to the non-vanishing commutator~\eqref{chargecomm} are excitations with gapless dispersion relation, rather than states with zero energy, so that contributions from ground states or other spurious states of intrinsically zero energy are ruled out. Taking the limit of infinite volume, and~$\mod\to0$, it can be explicitly shown~\cite{Lange:1966zz} that such states keep on not contributing to~\eqref{chargecomm}.

So, the actual, admissible contributions come from the first two terms in~\eqref{nonrelstruct}. The first one is the analogous of the unique term of the relativistic case, with the exception that there $\tilde{\rho}_1$ is forced by Lorentz symmetry to be just a constant. The second term has a completely transverse structure, so~$\tilde{\rho}_2$ is left completely arbitrary. This gives rise to the possibility of non-linear dispersion relations for non-relativistic Goldstone modes\footnote{%
	As a curiosity, we point out that such term was excluded by Guralnik, Hagen, and Kibble in their review~\cite{Guralnik:1967zz}, by a not convincing argument relying on the divergence of the spatial current at infinity. The legitimacy of this term has been restored by Nielsen and Chadha, who indicated it as responsible for non-linear dispersion relations~\cite{Nielsen:1975hm}, even though without giving any proof, and without confuting Guralnik-Hagen-Kibble's argument, and not even mentioning their work.
}, as we will show now, by an argument based on the symmetry breaking Ward identity. 

In previous section we have seen that spontaneous symmetry breaking entails a contact term in the Ward identity~\eqref{WIssb}. Such Ward identity holds in absence as in presence of Lorentz invariance, as long as the invariance of the measure and the continuity equation~$\partial_{\mu}J^\mu=0$ are assured. 

In the relativistic case, the  Ward identity~\eqref{WIssb} implied the presence of the Goldstone massless pole~\eqref{GBpoleWI}, with the completely fixed relativistic dispersion relation, $\omega\big(\mod\big)=\mod$, which is linear with proportionality constant given by the speed of light~$c\equiv1$. Here, with an argument inspired by~\cite{Brauner:2007uw} but somehow original, we use the Ward identity to derive the more general structure of the dispersion relation of non-relativistic GBs. 

The non-vanishing contact term proportional to the vev~\eqref{WIssb} implies a pole at dispersion relation $\omega_{\vec{k}}$, with~$\omega_{\vec{k}}\to0$ as~$\mod\to0$, in the correlator
\begin{equation}
\Langle J^\mu(k)X_A(-k)\Rangle = \bfrac{iv}{-k^0+\omega_{\vec{k}}} \left(n^\mu\,A\big(\mod\big) + b^\mu\,B\big(\mod\big) \vphantom{\Big|}\right)\ ,
\end{equation}
where we have allowed for a completely general non-relativistic structure (which yet preserves spatial rotational invariance), with~$n^\mu=\big(1,\vec{k}\big/\mod\big)$, `normalized' light-like vector, and~$b^\mu$ the same as in~\eqref{nonrelcomm}. In this way~$A$ and~$B$ have the same mass dimensions. Then the Ward identity becomes an equation for~$A$ and~$B$, giving the dispersion relation. Namely,
\begin{align}
iv = k_\mu\Langle J^\mu X_A\Rangle &=
	\bfrac{iv}{-k^0+\omega_{\vec{k}}}\; \Big(\left(-k^0+\mod\right)A -k^0\,B\Big) = \nn
&=
	iv\,\left(A+B\right) 
		+iv\;\bfrac{\big(-\omega_{\vec{k}}+\mod\big)\,A -\omega_{\vec{k}}\,B}{-k^0+\omega_{\vec{k}}}\ . \label{WIonshell}
\end{align}
The realization of the Ward identity demands the term in~$A+B$ to be a constant for $\mod\to0$, and the term diverging at the location of the Goldstone pole to vanish. We can write
\begin{align*}
A\big(\mod\big) &= a_0 +a_1\mod +\ldots\ ;\\
	B\big(\mod\big) &= b_0 +\ldots\ .
\end{align*}
In this way, the first condition that we extract from~\eqref{WIonshell}, in order to recover the Ward identity, is~$a_0+b_0=1$. Then, the other condition is that the second term must vanish, giving
\begin{equation*}
\big(-\omega_{\vec{k}}+\mod\big)\,A -\omega_{\vec{k}}\,B =0\ .
\end{equation*}

It is easy to check that for $b_0\neq1$ a linear dispersion relation occurs at low momentum
\begin{equation}
\omega_{\vec{k}} = \left(1-b_0\right) \mod\ ,
\end{equation}
with the relativistic one being a special case, at~$b_0=0$. If instead $b_0=1$ (and $a_1\neq0$), we have
\begin{equation}
\omega_{\vec{k}} = a_1 \veckk\ ,
\end{equation}
that is a quadratic dispersion relation. We will see in the following that such quadratic dispersion relation can be connected to a non-zero vacuum expectation value of a charge density, event that clearly can only occur when Lorentz invariance is violated.

As a final remark, we point out that throughout this discussion we have assumed the dispersion relation to be analytic in the momentum, as a consequence of the analyticity of the Fourier transform. This is guaranteed if there are no long-range interactions~\cite{Nielsen:1975hm}. However, Goldstone modes can occur also in the presence of long-range interactions (as long as the divergence of the spatial current is assured to vanish at infinity), and can exhibit non-analytic dispersion relation, for instance with fractional powers of momentum (see Section~6 of Brauner's review~\cite{Brauner:2010wm} for concrete examples).

Another main issue of the non-relativistic extension of Goldstone theorem concerns the number of Goldstone bosons in relation to the number of broken generators. In the relativistic case we have (at least\footnote{%
	The non-perturbative proof of Goldstone theorem presented here does not rule out the possibility that Goldstone modes may be more than broken generators. However, in all known example the strict equality holds, and proofs of Goldstone theorem based on perturbation theory and effective potential (see section~19.2 of Weinberg's book~\cite{Weinberg:1996kr}) show that there is a one-to-one correspondence between number of Goldstone modes and number of independent broken generators. Furthermore, the non-relativistic counting of~\eqref{nonrelcount} reduces to the equality between numbers of broken generators and Goldstone bosons for the relativistic sub-case $\mathrm{rank}\rho=0$.
}) one Goldstone particle for each broken generator. Already in early times Nielsen and Chadha~\cite{Nielsen:1975hm} showed that, in lack of Lorentz invariance, the possibility of dispersion relations proportional to even powers of momentum demands a modification of the counting rule, resulting in the inequality
\begin{equation}	\label{NielsenChadha}
n_{I}+2n_{II}\geq n_{BG}\ ,
\end{equation}
where ~$n_{BG}$ is the number of broken generators, and~$n_I$ and $n_{II}$ are the numbers of type~I and type~II Goldstone modes, with type~I having odd-power, and type~II even-power dispersion relations. Of course, relativistic Goldstone boson are all necessarily of type~I, giving the inequality~$n_{GB}\geq n_{BG}$, with $n_{GB}$ being the number of Goldstone bosons.

More than twenty-five years later~\cite{Schafer:2001bq} the missing counting was connected to non-vanishing vacuum expectation value for the charge operators. More precisely, the statement of~\cite{Schafer:2001bq} was that if 
\begin{equation}
\Langle\big[Q^a,Q^b\big]\Rangle = 0
\end{equation}
for all broken generators, then~$n_{GB}\geq n_{BG}$, as in the relativistic case. Of course, Lorentz invariance excludes the possibility of a non-zero vev for a charge, that is defined in term of the \emph{temporal} component of a current. Actually, the original statement of~\cite{Schafer:2001bq} was given as an equality, which does not necessary follows from their argument, and also their reasoning presented some inaccuracies. Anyway, the statement, in the form of an inequality, is correct, and the proof can be refined, as you can see for instance in section~5.3 of Brauner's review~\cite{Brauner:2010wm}.

The connection between charge density taking a vev and quadratic dispersion relation was already pointed-out by means of low-energy effective Lagrangian for non-relativistic Goldstones~\cite{Leutwyler:1993gf}. However, a precise counting rule has been established only relatively recently, again thanks to a low-energy effective Lagrangian approach with some assumed properties, in a series of papers~\cite{Brauner:2007uw,Watanabe:2011ec,Watanabe:2012hr,Hidaka:2012ym,Watanabe:2014fva}, whose conclusions summarize as follows:
\begin{equation}	\label{nonrelcount}
n_A=n_{BG}-\mathrm{rank}\rho\ , \quad 
	n_B = \bfrac12\mathrm{rank}\rho\ , \quad
		n_A+n_B= n_{GB}\ , \quad
			n_A+2n_B=n_{BG}\ ,
\end{equation}
where $n_A$ and $n_B$ are the number of type~A and type~B Goldstones respectively, and
\begin{equation}
i\rho_{ab} = \Langle\big[Q_a,J_b^0\big]\Rangle\ .
\end{equation}
The distinction of type~A and~B, rather than type~I and~II, is based on the counting rather than on dispersion relation: a type~A GB corresponds to a unique broken generator, whereas a type~B GB corresponds to a pair of canonically conjugated broken generators. Nonetheless, type~A GBs typically have linear dispersion relation, while type~B GBs have quadratic dispersion relation, even if there are exceptions, namely when the breaking of spacetime symmetries is involved~(see sec.~VI.A of~\cite{Watanabe:2014fva}).

In next section we will briefly analyze a basic field theory model where Lorentz invariance is explicitly broken by a chemical potential, and both type~A and type~B Goldstone bosons are encountered.

\bigskip

Along this first section, we have seen in a general fashion how Ward identities for symmetry breaking are a strongly universal feature, which applies to non-relativistic field theories in the same way as it does for relativistic ones. Moreover, we have pointed out how contact terms in the Ward identities are connected to the presence of massless Goldstone modes in the spectrum, and we have explored what information they can unveil over the dispersion relations of such gapless modes. 

In section~3, we will see how to retrieve the non-perturbative structure of Ward identities for symmetry breaking in holography. Before, let us introduce, in the next section, the holographic correspondence, and the main computational tool that will be adopted in section~3, \emph{i.e.}~holographic renormalization.

\clearpage{\pagestyle{empty}\cleardoublepage}
\section{Holography from a boundary-oriented perspective}
			
The duality between quantum, strongly-coupled field theories and classical, weekly-coupled gravitational theories has been originally conjectured~\cite{Maldacena:1997re} and affirmed between asymptotically Anti de Sitter (AdS) space-times and conformal field theories in flat space. Yet the holographic principle seems to hold in a much broader variety of different cases, as for instance in the case, relevant for us, of non-relativistic bulk geometries. As Jared Kaplan affirms in his lecture notes on AdS/CFT correspondence~\cite{Kaplan:2016}: ``AdS/CFT is many things to many people''. The gauge/gravity duality is then even more things for at least as many people, so let us say what it is to us.\footnote{%
	Giving a review on holography is far beyond the purpose of these notes. For complete and pedagogical introductions we refer to the copious existing material, as for instance~\cite{Klebanov:2000me,DHoker:2002nbb,Maldacena:2003nj,Nastase:2007kj,Polchinski:2010hw,Ramallo:2013bua,Penedones:2016voo}.}

Let us start from the original concrete realization of AdS/CFT~\cite{Maldacena:1997re}. Maldacena considered $N$ coincident D$3$-branes in type IIB string theory, where the low-energy limit provides two alternative descriptions. On the one hand, in perturbative string theory at low-energy, where only massless excitations are important, massless close strings have irrelevant interactions and decouple, while massless open strings ending on the D$3$-branes have dimensionless coupling and survive. The resulting theory of interacting SU$(N)$ gauge fields living on the four-dimensional (flat) world-volume of the $N$~D$3$-branes is $\mathcal{N}=4$ SU$(N)$ super-Yang-Mills (SYM), with Yang-Mills coupling $g_\mathrm{YM}$ given in terms of the string coupling~$g_s$, by the identification~$g_\mathrm{YM}^2\sim g_s$. On the other hand, D$3$-branes are black brane solutions of supergravity (low-energy classical limit of string theory). In this picture the low-energy limit selects the near-horizon region, and the near-horizon geometry of D$3$-branes in 10-dimensional type~IIB supergravity is given by AdS$_5\times S^5$, where $S^5$ is the five-sphere and the AdS radius~$L$ is given by~$L^4\sim g_sN\,l_s^4$, where $l_s$ is the string length. 

So, we have taken the same limit of a single theory, and obtained two apparently very different theories, which nonetheless should be equivalent: $\mathcal{N}=4$ SU$(N)$ super-Yang-Mills on flat space on one side, and supergravity on~AdS$_5\times S^5$ on the other. Let us first remark that the symmetries match on the two sides: the group of isometries of AdS$_5$ coincide with the conformal group in one less dimension,~SO$(4,2)$ (and this is true for any dimensions),  the symmetry of the five-sphere, SO$(6)$, matches the SU$(4)$ $R$-symmetry of $\mathcal{N}=4$ SYM, and supersymmetries are the same on both sides.

However, the two approximations are valid in opposite regimes, which is the essential property of a duality. Indeed, the perturbative description holds when perturbation theory applies, that is for $g_sN\ll1$. On the other hand supergravity, \emph{i.e.} the classical approximation, is reliable if the quantum corrections are negligible, so~$g_s\to0$, and if the AdS curvature radius~$L$ is larger than the string length~$l_s$, which is true for~$g_sN\gg1$, which implies~$N\to\infty$. 

Let us summarize by introducing the 't~Hooft coupling~$\lambda$~\cite{tHooft:1973alw}, so that
\begin{equation}	\label{tHooftcoupling}
\frac{L^4}{l_s^4}\sim g_sN \sim g_\mathrm{YM}^2N \equiv \lambda \ .
\end{equation}
If in SU$(N)$ Yang-Mills theory one takes $N$ to be large, but $g_\mathrm{YM}$ small enough so that $\lambda\ll1$, perturbation theory applies, and 't~Hooft showed~\cite{tHooft:1973alw} that at large~$N$ quantum loop diagrams organize according to the topology, with decreasing powers of~$N$ as the genus of the diagram increases. In this way, planar diagrams are dominant in the large~$N$ limit (which for this reason is also called planar limit).

In conclusion, if we can commute the low-energy limit with the fact of moving continuously between the weak-coupling regime~$\lambda\ll1$ and the strong-coupling regime~$\lambda\gg1$, then the duality is established. This last assumption is assumed to be true, but it is very hard to be proved or disproved, precisely for the fact that in the regime where one side of the duality is controllable, the other is hardly tractable, and vice-versa. At the same time, such duality is a very powerful and promising tool, precisely for the fact that when one of the side is obscure, the other is generically under control. 

Anyway, many non-trivial checks of the duality have been performed, in particular on supersymmetry-protected quantities, which can be computed on both sides~\cite{Aharony:1999ti}, but also by a large variety of constructions, which suggests that the duality holds beyond its original AdS/CFT framework. Actually, in the perspective of our discussion, we precisely wish it to extend to the broadest possible class of theories (namely non-relativistic space-time, non-conformal field theories, and so on). On the other hand, since we are mainly interested in using holography to study quantum field theories at strong coupling (in this sense, a boundary-oriented perspective), we are happy with the classical approximation for bulk gravity (which for our purposes will actually be even non-dynamical). 

Letting the quantum completion of our classical gravity be string theory as well as whatever other theory of quantum gravity, for the aims of these notes we assume
\begin{equation}
\frac{L^4}{L_{Pl}^4} \gg 1\ ,
\end{equation}
meaning that the size~$L$ of our bulk space-time is large with respect to the Planck length~$L_{Pl}$, that is the scale where quantum gravitational effects become relevant, so that the curvature is weak and the classical description of gravity holds. In turns, this implies that the dual QFT contains a large number of degrees of freedom.

In this range of validity, a weaker/broader holographic principle can be formulated, which is embodied by the following statement~\cite{Witten:1998qj}: the on-shell partition function of a classical gravitational theory in a given \mbox{$(d\!+\!1)$-di}mensional space-time~$\mathcal{M}$ is equivalent to the (off-shell) generating functional for correlation functions of a gauge field theory with no gravity on the boundary~$\partial\mathcal{M}$. In formul\ae:
\begin{align}	\label{HoloPrinciple}
e^{iS_\mathrm{cl}^\mathrm{grav}} \equiv 
	\exp\Big[i\int_{\mathcal{M}}\mathrm{d}^{d+1}\!x\, \g\; \mathcal{L}_{\mathrm{cl}}[\varphi]\Big|_{\raisebox{1.5pt}{\scriptsize{on-shell}}}\Big] &= \\
=
	\Big\langle\exp\Big[ i\int_{\partial\mathcal{M}}\mathrm{d}^d\!x\, \gh\; \varphi_0\,\mathcal{O}_{\varphi} \Big]\Big\rangle_{\!0} &
		\equiv \mathcal{Z}_\mathrm{QFT} ,	\nonumber
\end{align}
where $g$ is the determinant of the bulk metric, $\hat{g}$ the one of the corresponding induced metric on the boundary, $\varphi$ is a generic classical bulk field, $\varphi_0$ is its asymptotic boundary value, and~$\mathcal{O}_{\varphi}$ is the boundary operator defined in this way as dual to the bulk field~$\varphi$. Notice that the left-hand side of this equation explicitly relies on a weak-coupling local Lagrangian formulation, whereas the right-hand side is the generating functional for correlation function in a general QFT, which is valid regardless of the strength of the coupling. 

The actual meaning of the left-hand side will be made clear in the following, but it is already evident from the right-hand side, that equation~\eqref{HoloPrinciple} furnishes a prescription for computing QFT $n$-point correlation functions. Defining $\mathcal{W}_\mathrm{QFT}=\ln\mathcal{Z}_\mathrm{QFT}$, we have the following formula for \emph{connected} $n$-point correlation functions
\begin{align}
\Langle \CO_{\!A_1\!}(x_1)\cdots\CO_{\!A_n\!}(x_n)\Rangle &\equiv 
	(-i)^n\left.\bfrac{\delta^n\;\mathcal{W}_\mathrm{QFT}[\varphi_0]\hspace{1.1cm}}{\delta\varphi_0^{\!A_1}(x_1)\cdots\delta\varphi_0^{\!A_n}(x_n)}\right|_{\raisebox{4pt}{$\scriptstyle\varphi_0=0$}} 
		\label{corrfuncts}\\
&=
	(-i)^n\left.\frac{\delta^n\;iS_\mathrm{cl}^\mathrm{grav}\hspace{1.7cm}}{\delta\varphi_0^{\!A_1}(x_1)\cdots\delta\varphi_0^{\!A_n}(x_n)}\right|_{\raisebox{4pt}{$\scriptstyle\varphi_0=0$}}
		\label{holocorrfuncts}
\end{align} 
where we have allowed for various boundary operators~$\CO_{\!A}$, dual to various bulk fields~$\varphi^{\!A}$.

Furthermore, we remark that the boundary dual operator has not to be a scalar, it can be any sort of composite QFT expressions, however, since it couples to the source~$\varphi_0$ in the generating functional~$\mathcal{Z}_\mathrm{QFT}$, it must have the same symmetry properties and quantum numbers as the corresponding bulk field~$\varphi$. In particular, for a conserved current~$J^\mu$ we have
\begin{equation}	\label{currentsource}
\exp\Big[ i\int_{\partial\mathcal{M}}\mathrm{d}^d\!x\, \gh\; A_{0\mu}J^\mu \Big]\ .
\end{equation}
As already discussed in Section~\ref{explicitWI} around eq.~\eqref{totlocvar}, the gauge invariance of the measure in the generating functional entails that the source of a conserved current undergoes gauge transformations. Thus, in holography this implies that the dual field of the conserved current of a global symmetry on the boundary  is given by an actual gauge field in the bulk.

Let us now focus on the left-hand side of the foundational equation~\eqref{HoloPrinciple}: the bulk action should be put on-shell by means of the bulk equations of motion, and the sources for dual operators identified. Then, suitable boundary conditions, and the capability to solve the equations of motion, allow to express the boundary action in terms of the sources only. However, the operation of reduction on the boundary may produce diverging terms. Such divergences can be removed in a consistent way, identifying at the same time the sources for dual operators, by a procedure called holographic renormalization, by analogy with its field theoretical counterpart. Next section is devoted to illustrate such procedure in detail.

\subsection{Holographic renormalization}
\label{HoloRen}

The idea of holographic renormalization was first proposed in~\cite{Henningson:1998gx,Henningson:1998ey}, and then developed in a systematic way in~\cite{deHaro:2000vlm,Bianchi:2001de,Bianchi:2001kw}. Here we illustrate it through the minimal example of a real scalar field~$\varphi$ of mass $m_{\varphi}$ on an~AdS$_{d+1}$ fixed background, following an approach similar to~\cite{Skenderis:2002wp}.

We thus consider the following bulk action,
\begin{equation}	\label{Srealscalar}
S= -\bfrac12\intd{d+1}\,{\g}\ \Big( g^{mn}\partial_m\varphi\partial_n\varphi +m_{\varphi}^2\varphi^2 \Big)\ ,
\end{equation}
where Latin letters indicate indices in the $(d+1)$-dimensional bulk, while Greek letters indicate indices in the $d$-dimensional boundary, which are contracted in the usual way with mostly-plus Minkowski metric\footnote{
	We choose to adopt from the beginning the `physical' Lorentzian signature, rather than the Euclidean one, even if the holographic renormalization and the computation of correlation functions require careful treatments~\cite{Skenderis:2008dh,Skenderis:2008dg}, related to the causal~$i\epsilon$ prescription for (time-ordered, retarded, advanced) correlation functions in QFT. However, for the setups presented in this work, which involve no temperature and no time-evolution, the actual Lorentzian computation would be uselessly complicated and equivalent to the Euclidean one with a final suitable analytic continuation to real time (Wick rotation), and indeed it is what we will be implicitly doing all the time.
}, and where the AdS metric in the Poincar\'e patch is defined by
\begin{equation}	\label{AdSmetric}
{\de{s}}^2 = \frac{1}{r^2}\left(\de{r}^2+\de{x}^\mu\de{x}_\mu\right)\ ,\qquad \text{so that }\quad
	g^{rr}=r^2\ , \quad g^{\mu\nu}=r^2\eta^{\mu\nu}\ ,
\end{equation}
where we have set the AdS radius~$L\equiv1$. We call~$r$ holographic coordinate or radial coordinate, and the boundary is located at~$r=0$.

The variational principle gives the following equation of motion
\begin{equation}	\label{eomvarphi}
0= -\bfrac{1}{\g}\partial_m\left(\!\g\bfrac{\delta\mathcal{L}}{\delta\partial_m\varphi}\right) -\bfrac{\delta\mathcal{L}}{\delta\varphi} =
	r^{d+1}\partial_r\Big(r^{-d+1}\partial_r\varphi\Big) +r^2\Box\varphi -m_\varphi^2\varphi\ ,
\end{equation}
where with the box we indicate the d'Alembertian in Minkowski space, $\Box=\partial_\mu\partial^\mu$. Exact analytic solutions to this equation are given in terms of Bessel functions, as we will see in Section~\ref{HoloRen2p}. For the moment, we just need the asymptotic behavior, which can be easily determined by solving the equation order by order in the near-boundary expansion, obtaining~\cite{Klebanov:1999tb,Witten:1998qj}
\begin{align}
\varphi= r^{d-\Delta}\left(\varphi_0 +r^2\,\varphi_1 +\;\ldots\; \right) +r^\Delta\left(\tilde{\varphi}_0 +r^2\,\tilde{\varphi}_1 +\;\ldots\; \right) &\ ,	\label{varphiasympt}\\
	\text{with }\ 	\Delta=\bfrac{d}{2}+\nud	&\ .	\label{Delta}
\end{align}
This power expansion holds whenever $2\Delta-d$ is different from zero, and is not an even integer, so for~$\Delta\neq\frac{d}{2}+n$, with~$n$ non-negative integer. Otherwise, the two parts of~\eqref{varphiasympt} mix and logarithmic terms arise, as it can be checked from the equation of motion~\eqref{eomvarphi} itself. The presence of logarithms does not represent any insurmountable obstacle, nonetheless we avoid them for the moment in order to keep the discussion simpler, and we postpone their treatment to Section~\ref{HoloRenlog}.

The value $\Delta=\tfrac{d}{2}$ is a minimal value for $\Delta$, and it corresponds to the minimal value that the scalar mass can get for the argument of the square root in the definition of~$\Delta$ to be non-negative, that is
\begin{equation}
m_\varphi^2 = -\frac{d^2}{4}\ .
\end{equation}
This defines the BF (Breitenlohner-Freedman) bound for the AdS scalar mass~\cite{Breitenlohner:1982jf}. Masses below this value correspond to tachyons in AdS. Nonetheless, AdS squared mass can get negative values, with still positive energy, since AdS curvature compensates with a positive contribution to the energy.

If we identify the asymptotic leading term~$\varphi_0$ as the source for the dual boundary operator~$\CO_\varphi$~\eqref{HoloPrinciple}, then~$\Delta$ (that is the asymptotic exponent of the sub-leading mode~$\tilde{\varphi}_0$) turns out to be the scaling dimension of~$\CO_\varphi$, so that we will call~$\tilde{\varphi}_0$ the `response'. Hence, the BF bound constitutes a lower bound for the dimension of the boundary operator: $\Delta\geq\frac{d}{2}$. Actually, in a conformal field theory there is a lower bound for the dimension of scalar operators, given by unitarity~\cite{Mack:1975je}:
\begin{equation}	\label{unitarybound}
	\Delta>\frac{d}{2}-1\ .
\end{equation}
However, the BF bound is above the unitarity bound, and there are no reasons from the boundary point of view for the dimension of~$\CO_\varphi$ not to be allowed to go down to the unitarity bound. And indeed values of the scaling dimension between the unitarity bound and the BF bound can be accessed by making~$\tilde{\varphi}_0$ the source, so that the the dual scalar operator is of dimension~$d-\Delta$ (\emph{i.e.}~$\varphi_0$ is the response).

Breitenlohner and Freedman themselves~\cite{Breitenlohner:1982jf}, demanding finiteness of the AdS action and positive definiteness of the energy, actually showed that for $m_\varphi^2>-\frac{d^2}{4}+1$, that is $\Delta>\frac{d}{2}+1$, only one choice of boundary condition ($\varphi_0$ is the source) is acceptable. Instead, for
\begin{equation}	\label{altquant}
	-\frac{d^2}{4}<m^2_\varphi<-\frac{d^2}{4}+1
\end{equation}
two alternative AdS-invariant quantizations are admissible. For $\varphi_0$ as source, we have an operator of dimensions~$\Delta$, with~$\frac{d}{2}<\Delta<\frac{d}{2}+1$; for~$\tilde{\varphi}_0$ as source, we have an operator of dimensions~$\tilde{\Delta}\equiv{d-\Delta}$, with~$\frac{d}{2}-1<\tilde{\Delta}<\frac{d}{2}$, filling in this way the window of dimensions between the BF bound and the unitarity bound~\cite{Klebanov:1999tb,Minces:1999eg,Minces:2001zy,Rivelles:2003ge}.

We will see in detail how to set these different boundary conditions, and how holographic renormalization, besides removing divergences, actually allows to fix which are the sources, and consequently select the quantization.

Now, let us go back to the action~\eqref{Srealscalar}, and try to reduce it to the boundary, in order to be able to apply our holographic principle~\eqref{HoloPrinciple}. We first integrate by parts, and use the equation of motion to put the action on shell:
\begin{align}
S= -\bfrac12\intd{d+1}\,&\Big[
	\partial_r\Big(r^{-d+1}\varphi\partial_r\varphi\Big)
		-\varphi\partial_r\big(r^{-d+1}\partial_r\varphi\big) -r^{-d+1}\varphi\Box\varphi +r^{-d-1}m_{\varphi}^2\varphi^2 \Big] \nn
= -\bfrac12\intd{d+1}\,&\phantom{\Big[}
	\partial_r\Big(r^{-d+1}\varphi\partial_r\varphi\Big)\ .
\end{align}
We are left with a total derivative in the radial coordinate, so the action can be reduced to a boundary term\footnote{%
	The on-shell action reduces to a purely boundary term because the considered action is quadratic. With cubic interactions, for instance, bulk terms would survive. However, as far as \emph{two-point} functions are concerned, the quadratic part of the action is all we need.
}, and expanded according to the asymptotic expression~\eqref{varphiasympt},
\begin{align}
S_{\mathrm{reg}}=\bfrac12&\intd[r=\epsilon\!]{d}\
	r^{-d}\:\varphi\,r\partial_r\varphi\ = 	\label{Sregvarphi}\\
 =\bfrac12&\intd[r=\epsilon\!]{d}\,
	\left(r^{d-2\Delta}\left(d-\Delta\right)\varphi_0 +r^{d-2\Delta+2}\,2(d-\Delta+1)\,\varphi_1 +\;\ldots\; +d\,\tilde{\varphi}_0\right)\varphi_0\ ,	\nonumber
\end{align}
where we used a small~$\epsilon$ as regulator for the potentially divergent terms, and the dots represent sub-leading terms of the order~$O\big(r^{d-2\Delta+4}\big)$. We can see that the leading term is always divergent, the minimal value of $\Delta$ being~$\tfrac{d}{2}$, that is at the BF bound, where the divergence is logarithmic, as we have already pointed out.

Let us focus first on the most dangerous term. We want to remove it by a boundary counterterm, which should not drastically modify our bulk theory. Then, we demand the counterterm to be \emph{local} and invariant under all the symmetries induced on the boundary by the bulk action (gauge invariance, space-time symmetries and internal symmetries). For the specific case of the action~\eqref{Srealscalar}, which possesses Poincar\'e symmetry and a global~$\mathbb{Z}_2$ invariance, the only terms we can write with the field content of the bulk action, \emph{i.e.} the real scalar~$\varphi$, are a kinetic term or a mass term. We will see that the former actually leads to alternative quantization, but we consider first the latter, namely
\begin{align}
S_{\mathrm{ct}}^{(0)} &= 
	-\bfrac12\intd{d}\: \gh\; \varphi^2 = 	\label{Sct0varphi}\\
& =	
	-\bfrac12\intd{d}\;
		\left(r^{d-2\Delta}\,\varphi_0 +r^{d-2\Delta+2}\,2\varphi_1 +\;\ldots\; +2\,\tilde{\varphi}_0\right)\varphi_0\ ,	\nonumber
\end{align}
where $\hat{g}_{\mu\nu}$ is the $d$-dimensional metric induced on the boundary by the bulk \mbox{$(d+1)$-di}mensional metric~$g_{mn}$, which in the present case is just the flat Min\-kowski metric rescaled by $r^2$: $\hat{g}_{\mu\nu}=r^{-2}\eta_{\mu\nu}$.

This counterterm serves to cancel the leading divergence in the regularized action~\eqref{Sregvarphi}, but leaves sub-leading divergences whenever~$\Delta\geq\frac{d}{2}+1$,
\begin{equation}	\label{SregSct0varphi}
S_\mathrm{reg}+\left(d-\Delta\right)S_{\mathrm{ct}}^{(0)} = 
	\bfrac12\intd{d}\; 
		\left(r^{d-2\Delta+2}\,2\varphi_1 +\;\ldots\; +\left(2\Delta-d\right)\tilde{\varphi}_0\right)\varphi_0\ .
\end{equation}
However, the coefficients of sub-leading terms~$\varphi_n$ can always be related to the coefficient of the leading term~$\varphi_0$, by means of the equation of motion. This traces the road to write the additional counterterms we may need.

For instance for the term proportional to~$\varphi_1$, which by the equations of motion is given by
\begin{equation}	\label{varphi1BOXvarphi0}
\varphi_1=\bfrac{1}{2(2\Delta-d-2)}\,\Box\varphi_0\ ,
\end{equation}
we have
\begin{align}
S_\mathrm{ct}^{(1)} &= 
	\bfrac12\intd{d}\: \gh\; \hat{g}^{\mu\nu}\varphi\partial_\mu\partial_\nu\varphi = 
		\bfrac12\intd{d}\; \left(r^{d-2\Delta+2}\,\varphi_0\Box\varphi_0 +\;\ldots\;\right) = \nn
&= 
	\bfrac12\left(2\Delta-d-2\right)\intd{d}\; \left(r^{d-2\Delta+2}\,2\varphi_0\varphi_1 +\;\ldots\;\right)\ , 	\label{Sct1}
\end{align}
which, dressed with the suitable numerical factor, is indeed able to cancel the leading surviving divergence in~\eqref{SregSct0varphi}. The dots here represent sub-leading terms of order~$O\big(r^{d-2\Delta+4}\big)$. They can still be divergent, but again we can use the equation of motion, which actually give, for any~$n$,
\begin{equation}
\varphi_n = \frac{1}{2n(2\Delta-d-2n)}\,\Box\varphi_{n-1} =
	\frac{\Gamma\big[\Delta-\tfrac{d}{2}-n\big]}{4^nn!\,\Gamma\big[\Delta-\tfrac{d}{2}\big]}\, \Box^n\varphi_0\ ,
\end{equation}
as it can be easily verified by induction. Thus, the sub-leading divergences can be removed, order by order, through a counterterm~$S^{(n)}_\mathrm{ct}$, similar to~\eqref{Sct1}, but with~$n$ powers of d'Alembert operator, until no divergent terms are left.

However, for the purpose of computing correlators and Ward identities, we do not even need to do that. Indeed, with the exception of the special case where the scalar mass saturates the BF bound\footnote{
	At the BF bound the presence of a logarithm in the leading term affects the definition of the source, forcing the introduction of a scale and so partially violating conformal invariance~\cite{Witten:2001ua,Marolf:2006nd}.
}, the term proportional to~$\varphi_0\tilde{\varphi}_0$ (\emph{i.e.}~to the two independent leading modes of the solution~\eqref{varphiasympt}, which will be identified with the source and the vev of the dual operator) is always finite in the leading counterterm~$S_\mathrm{ct}^{(0)}$~\eqref{Sct0varphi}, and always vanishing in the rest of sub-leading counterterms~$S_\mathrm{ct}^{(n)}$\footnote{%
	This fact has been shown more systematically using a different renormalization technique based on dimensional regularization~\cite{Bzowski:2016kni}.
}. Thus, precisely this term, bi-linear in the leading and sub-leading, yields the main part of the renormalized action
\begin{equation}	\label{Srenvarphi}
S_\mathrm{ren}= \left(\Delta-\bfrac{d}{2}\right)\intd{d}\; \varphi_0\tilde{\varphi}_0\ .
\end{equation}
When logarithmic divergences are present, this renormalized action gets modified by scheme-dependent finite pieces, as we will see in Section~\ref{HoloRenlog}.

Now, we affirm that, in this renormalization scheme~\eqref{Sct0varphi}, $\varphi_0$ is the source for the dual boundary operator~$\CO_\varphi$, which has dimension~$\Delta$, with~$\Delta>\bfrac{d}{2}$ by the definition~\eqref{Delta}.\footnote{
	In the so-called alternative quantization scheme, the renormalized action will be changed by a sign (see eq.~\eqref{Srenvarphitil}) and the source would be~$\varphi$, sourcing a boundary operator of dimension~$\tilde{\Delta}=d-\Delta$, with~$-\bfrac{d}{2}<\tilde{\Delta}<\bfrac{d}{2}$, as we will see in the following.}

In the path integral formulation, the source is a field that is coupled to a given operator of the theory, and whose variation should be set to zero in order to preserve the variational principle (invariance of the action). In this way, the source is dual to the associated operator, and allows to generate the correlation functions for that operator.

So, in order to verify that $\varphi_0$ is actually the source, we have to check the variational principle on the renormalized action~\eqref{Srenvarphi}. The correct variation of the renormalized action is not the na\"ive variation of expression~\eqref{Srenvarphi}, is rather the variation of the bulk action~\eqref{Srealscalar}, put on-shell and reduced to the boundary, and then renormalized by the variation of employed counterterms, which removes divergent pieces. Basically we have to repeat the renormalization procedure on the variation~$\delta{S}$, so in practice the variation of the renormalized action is rather the renormalization of the varied action.

For the on-shell variation of the bulk action we have
\begin{align}
\vardel{S} &=
	-\intd{d+1}\,{\g}\ \Big( g^{mn}\partial_m\varphi\partial_n\delta\varphi +m_{\varphi}^2\varphi\delta\varphi \Big) =
		\intd[r=\epsilon\!]{d}\ r^{-d}\:\delta\varphi\,r\partial_r\varphi =\nn
&=
	\intd[r=\epsilon\!]{d} \left[(d-\Delta)r^{d-2\Delta}\,\varphi_0\delta\varphi_0 +\;\ldots\; +\Delta\,\tilde{\varphi}_0\delta\varphi_0 +(d-\Delta)\varphi_0\delta\tilde{\varphi}_0\right]\ .
	\label{varSvarphi}
\end{align}
On the other hand, the variation of counterterm~\eqref{Sct0varphi} yields
\begin{equation*}
\vardel{S}_\mathrm{ct}^{(0)} =
	-\intd{d}\: \gh\; \varphi\delta\varphi =
		-\intd{d}\;\left[r^{d-2\Delta}\,\varphi_0\delta\varphi_0 +\;\ldots\; +\tilde{\varphi}_0\delta\varphi_0 +\varphi_0\delta\tilde{\varphi}_0\right]\ ,
\end{equation*}
which, with the same numerical factor as in~\eqref{SregSct0varphi}, removes the leading divergence. Possible additional divergences can be removed by the variation of the respective counterterms, which yet do not affect the finite terms. Thus, we finally have
\begin{equation}
\vardel{S}_\mathrm{ren}=\vardel{S}+\left(d-\Delta\right)\vardel{S}_\mathrm{ct}^{(0)} = \left(2\Delta-d\right)\intd{d}\ \tilde{\varphi}_0\,\delta\varphi_0\ ,
\end{equation}
where only the variation of~$\phi_0$ appears. The variational principle for the boundary action imposes~$\delta\varphi_0=0$, so that~$\varphi_0$ is correctly identified as the source for the dual operator.

Finally, we remark that, once $\varphi_0$ is established as the source, the response~$\tilde{\varphi}_0$ can be related to the vacuum expectation value of the dual operator~$\CO_\varphi$. Indeed, applying~\eqref{holocorrfuncts} on the renormalized action~\eqref{Srenvarphi},
\begin{equation}	\label{1point}
\Langle\CO_\varphi\Rangle = \left.\bfrac{\delta S_\mathrm{ren}}{\delta\varphi_0}\right|_{\raisebox{3pt}{$\scriptstyle\varphi_0=0$}}=
	\left(2\Delta-d\right)\left.\tilde{\varphi}_0\right|_{\varphi_0=0}\ .
\end{equation}
For these reason the response is usually referred to as the `vev'. A fixed (non-vanishing with the source) value for the coefficient of the sub-leading mode, then returns a non-vanishing vacuum expectation value for the dual operator in the boundary theory. This is relevant when we want to describe spontaneous symmetry breaking in the boundary field theory.

We will now show that a different counterterm can cancel the same divergence as~\eqref{Sct0varphi}, yet giving the opposite quantization, where~$\tilde{\varphi}_0$ is the source and~$\varphi_0$ the response. We recall that we are in the window~\eqref{altquant}, where double quantization is possible, so $\frac{d}{2}<\Delta<\frac{d}{2}+1$, and for this range of values of~$\Delta$ the leading divergence is the unique divergence in the on-shell action~\eqref{Sregvarphi}~\cite{Minces:2001zy,Rivelles:2003ge}.

We consider the following boundary term,
\begin{align}
\tilde{S}_\mathrm{ct}^{(0)} &= 
	-\bfrac12\intd{d}\: \gh\; \big(r\partial_r\varphi\big)^2\ =	\label{tilSct0varphi}\\
& =
	-\bfrac{d-\Delta}{2}\intd{d}\; \left((d-\Delta)r^{d-2\Delta}\,\varphi_0 +2\Delta\,\tilde{\varphi}_0\right)\varphi_0\ .	\nonumber
\end{align} 
We point out that this counterterm is made out of the square of the conjugate momentum of~$\varphi$ in the holographic coordinate, and it is actually related to counterterm~\eqref{Sct0varphi}, which is a square in~$\varphi$, by a Legendre transform\footnote{%
	The fact that the holographic generating functional in alternative quantization is related to the holographic generating functional in standard quantization by a Legendre transform was first suggested in~\cite{Klebanov:1999tb}.
}, namely
\begin{equation}	\label{Legendrevarphi}
\intd{d}\varphi\Pi_\varphi -\left(d-\Delta\right)S_\mathrm{ct}^{(0)} = \bfrac{1}{d-\Delta}\,\tilde{S}_\mathrm{ct}^{(0)}\ ,
\end{equation}
where~$\Pi_\varphi\equiv r\partial_r\varphi$. This expression actually furnishes the suitable numerical factor for~\eqref{tilSct0varphi} to cancel the leading divergence in the regularized action~\eqref{Sregvarphi}, yielding
\begin{equation}	\label{Srenvarphitil}
\tilde{S}_\mathrm{ren}= S_\mathrm{reg}+\bfrac{1}{d-\Delta}\,\tilde{S}_{\mathrm{ct}}^{(0)} =
	-\left(\Delta-\bfrac{d}{2}\right)\intd{d}\; \varphi_0\tilde{\varphi}_0 = \left(\tilde{\Delta}-\bfrac{d}{2}\right)\intd{d}\; \varphi_0\tilde{\varphi}_0 \ .
\end{equation}
So, the renormalized action for a real scalar in alternative quantization is identical to the one in standard quantization~\eqref{Srenvarphi}, except for the opposite sign, which again corresponds to replacing~$\Delta$ by~$\tilde{\Delta}=d-\Delta$.

It is straightforward now to check the variational principle on this last renormalized action. From the variation of the bulk action~\eqref{varSvarphi} and the variation of counterterm~\eqref{tilSct0varphi}, we obtain
\begin{equation}
\vardel\tilde{S}_\mathrm{ren}=\vardel{S}+\bfrac{1}{d-\Delta}\,\vardel\tilde{S}_\mathrm{ct}^{(0)} = \left(d-2\Delta\right)\intd{d}\ \varphi_0\,\delta\tilde{\varphi}_0\ ,
\end{equation}
which indeed requires to set~$\delta\tilde{\varphi}_0$ to zero, so that now $\tilde{\varphi}_0$ is the source.

We are now ready to use the holographic principle~\eqref{HoloPrinciple} to explicitly compute two-point correlation functions for the dual operator~$\CO_\varphi$ of dimension~$\Delta$ (in standard quantization), and for~$\tilde{\CO}_\varphi$ of dimension~$\tilde{\Delta}$ (in alternative quantization).

\subsubsection{Two-point functions from the renormalized action}\label{HoloRen2p}

The two-point correlation function is by definition a \emph{non-local} function. Let us define, for conciseness,
\begin{equation}	\label{nonlocalcorr}
\Langle \CO_\varphi(x)\CO_\varphi(x\p)\Rangle \equiv -i f_\varphi(x-x\p)\ ,
\end{equation}
and then express the generating functional~\eqref{corrfuncts} at quadratic order in the sources,
\begin{equation*}
\mathcal{W}_\mathrm{QFT}[\varphi_0] = \frac{i}{2}\int\!\de{y}\,\de{y'}\,\varphi_0(y)f_\varphi(y-y')\varphi_0(y')\ ,
\end{equation*}
so that the formula~\eqref{corrfuncts} on this last expression correctly reproduces~\eqref{nonlocalcorr}. This suggests that, in the standardly quantized renormalized action~\eqref{Srenvarphi}, the response~$\tilde{\varphi}_0$ should be related to the source by a non-local function, namely
\begin{equation}	\label{nonlocfvarphi}
\left(2\Delta-d\right)\tilde{\varphi}_0(x) = f_\varphi(x-x\p)\varphi_0(x\p)\ .
\end{equation}

This fact is achieved by imposing `boundary'\footnote{%
	We apologize for the unfortunate recurrence of the word boundary, in this case to design the conditions at the extremes of a boundary value differential problem, so that we have \emph{boundary} boundary conditions and \emph{bulk} boundary conditions	
} conditions in the bulk. Indeed, with the renormalization procedure we have fixed boundary conditions on the boundary of AdS. In order to have a unique solution to the second-order equation~\eqref{eomvarphi}, we need to fix another boundary condition, on the opposite side, that is in the deep bulk~$r=\infty$. Precisely this latter will relate the source and the response of the near-boundary expansion, as we will see explicitly in a while for the example presented here.

Before, let us apply the formula~\eqref{holocorrfuncts} to the renormalized action in standard quantization~\eqref{Srenvarphi}, and use~\eqref{nonlocfvarphi} to obtain
\begin{equation}	\label{2pointformulastd}
\Langle \CO_\varphi(x)\CO_\varphi(x\p)\Rangle =
	-i\left.\bfrac{\delta^2S_\mathrm{ren}}{\delta\varphi_0\delta\varphi_0}\right|_{\raisebox{2.5pt}{$\scriptstyle\varphi_0=0$}} =
		-i\left(2\Delta-d\right)\frac{\delta\tilde{\varphi}_0(x\p)}{\delta\varphi_0(x)}=-if_\varphi(x-x\p)\ ,
\end{equation}
again consistently with our renaming~\eqref{nonlocalcorr}.

If we are in alternative quantization, from~\eqref{Srenvarphitil} we get
\begin{equation}	\label{2pointformulaalt}
\Langle \tilde{\CO}_\varphi(x)\tilde{\CO}_\varphi(x\p)\Rangle =
	-i\left.\bfrac{\delta^2\tilde{S}_\mathrm{ren}}{\delta\tilde{\varphi}_0\delta\tilde{\varphi}_0}\right|_{\raisebox{2.5pt}{$\scriptstyle\tilde{\varphi}_0=0$}} =
		+i\left(2\Delta-d\right)\frac{\delta\varphi_0(x\p)}{\delta\tilde{\varphi}_0(x)}=-i\tilde{f}_\varphi(x-x\p)\ ,
\end{equation}
where we have defined
\begin{equation} \label{nonlocfvarphitil}
-\left(2\Delta-d\right)\varphi_0(x)=\tilde{f}_\varphi(x-x\p)\tilde{\varphi}_0(x\p)\ . 
\end{equation}

Now we have to find a full solution to the equation of motion, in order to explicitly determine the non-local functions~$f_\varphi,\tilde{f}_\varphi$. Consider the equation of motion~\eqref{eomvarphi} in momentum space,
\begin{equation}	\label{eomvarphip}
r^{d+1}\partial_r\Big(r^{-d+1}\partial_r\varphi_k\Big) -k^2\varphi_k +\Delta(d-\Delta)\varphi_k =0\ ,
\end{equation}
where $\varphi_k$ is the boundary Fourier transform of the field, defined by
\begin{equation}
\varphi(r,x)=\int\bfrac{\de^d\!k}{(2\pi)^d}\ e^{ikx}\,\varphi_k(r)\ .
\end{equation}
Rescaling $\varphi_k(r)\equiv r^{\frac{d}{2}}\bar{\varphi}(k,r)$, and then performing the change of variable~$\rho=kr$,\footnote{\label{kreal}%
	Here by $k$ we rather mean~$\sqrt{k^2}$, which will assume to be real when we impose the bulk boundary condition here after. This corresponds to~$k^2>0$, rather than the other way round, as we would expect in our mostly-plus signature. However, the spectrum of the final correlator will be on the right side, \emph{i.e.}~for $k^2<0$. Actually it is a quite general feature of holography that the domain imposed by bulk regularity requirements and the region where the spectrum is located be complementary.
}
this equation reduces to the modified Bessel equation
\begin{equation}
\rho^2\,\bar{\varphi}\p\p(\rho) +\rho\,\bar{\varphi}\p(\rho) -\left(\rho^2+\big(\Delta-\tfrac{d}{2}\big)^2\right)\bar{\varphi}(\rho) =0\ ,
\end{equation}
whose solution is given by
\begin{equation}
\bar{\varphi}(\rho)= C_1\; \sfI[\nu;\,\rho] +C_2\;\sfK[\nu;\,\rho]\ ,
\end{equation}
where we have defined~$\nu=\Delta-\tfrac{d}{2}$.

As announced at the beginning of this section, we have to impose boundary conditions in the interior in order to get rid of one integration constant, the other one being solved by boundary conditions at the boundary. As you can check in your favorite reference on special functions, the asymptotic behavior of the modified Bessel function of the first kind~$\sfI[\nu;\rho]$ for~$\rho\to\infty$ is $\sfI\sim \rho^{-\bfrac12}e^\rho+e^{-\rho}$, while the asymptotic behavior of the modified Bessel function of the second kind~$\sfK[\nu;\rho]$ is~$\sfK\sim \rho^{-\bfrac12}e^{-\rho}$. We prefer our solution not to explode in the deep bulk, and rather be small, if we want bulk gravitational effects be negligible. If we assume $\rho$ real and positive (that means, as priorly discussed in footnote~\ref{kreal}, $k$ real and positive), the function~$\sfK$ vanishes at~$r=\infty$. 

So, we choose $C_1=0$, and we finally obtain
\begin{equation}	\label{solvarphi}
\varphi_k = C_2\,r^{\frac{d}{2}}\;\sfK[\Delta-\tfrac{d}{2};\,kr\big]\ .
\end{equation}
For $r\to0$, this solution correctly presents
\begin{equation*}
\varphi_k = 
	r^{d-\Delta}\ \frac{C_2}{2}\:\Gamma\Big[\Delta-\tfrac{d}{2}\Big]\left(\frac{k}{2}\right)^{\tfrac{d}{2}-\Delta} +\;\ldots\; +
		r^{\Delta}\ \frac{C_2}{2}\:\Gamma\Big[\tfrac{d}{2}-\Delta\Big]\left(\frac{k}{2}\right)^{\Delta-\tfrac{d}{2}} +\;\ldots\;\ ,
\end{equation*}
which, by comparison with the boundary expansion~\eqref{varphiasympt}, allows to determine
\begin{equation}
\varphi_0 = \frac{C_2}{2}\:\Gamma\Big[\Delta-\tfrac{d}{2}\Big]\left(\frac{k}{2}\right)^{\tfrac{d}{2}-\Delta}\ , \qquad
	\tilde{\varphi}_0= \frac{C_2}{2}\:\Gamma\Big[\tfrac{d}{2}-\Delta\Big]\left(\frac{k}{2}\right)^{\Delta-\tfrac{d}{2}}\ ,
\end{equation}
where, with a little abuse of notation, we indicate the Fourier transforms in the same way as their counterparts in position space. The reader can notice that so far we have not specified anything about which is the source and which is the response. The solution of the equation of motion is one and unique, giving the asymptotic expansion~\eqref{varphiasympt}. Everything about quantization, sources and responses is contained in the renormalization procedure, which selects either the prescription~\eqref{2pointformulastd} or~\eqref{2pointformulaalt}.

Since we are in momentum space, the relation~\eqref{nonlocfvarphi} becomes an algebraic relation, that is~$(2\Delta-d)\,\tilde{\varphi}_0=f_\varphi(k)\,\varphi_0$, so that in standard quantization, from~\eqref{2pointformulastd}, we have
\begin{equation*}
\Langle\CO_\varphi(k)\CO_\varphi(-k)\Rangle=-if_\varphi(k) = -i\left(2\Delta-d\right)\bfrac{\tilde{\varphi}_0}{\varphi_0} =
	-i\left(2\Delta-d\right)\bfrac{\Gamma\big[\tfrac{d}{2}-\Delta\big]}{\Gamma\big[\Delta-\tfrac{d}{2}\big]}\left(\bfrac{k}{2}\right)^{2\Delta-d}\ .
\end{equation*}
In order to obtain the correlator in position space, we have to Fourier transform. Using the formula
\begin{equation*}
\int\bfrac{\de^d\!k}{(2\pi)^d}\ k^n = \frac{2^n}{\pi^\frac{d}{2}}\bfrac{\Gamma\big[\frac{d+n}{2}\big]}{\Gamma\big[-\frac{n}{2}\big]}\; \bfrac{1}{|x|^{d+n}}\ ,
\end{equation*}
we obtain
\begin{equation}
\Langle\CO_\varphi(x)\CO_\varphi(0)\Rangle = -i\,\bfrac{2\Delta-d}{\pi^\frac{d}{2}}\bfrac{\Gamma\big[\Delta\big]}{\Gamma\big[\Delta-\tfrac{d}{2}\big]}\; \bfrac{1}{|x|^{2\Delta}}\ ,
\end{equation}
which is precisely the expression in conformal field theory for the two-point correlator of a primary operator of scaling dimension~$\Delta$, thus confirming our holographic dictionary for~$\Delta$.

On the other hand, in alternative quantization we obtain from~\eqref{2pointformulaalt}
\begin{align*}
\Langle\tilde{\CO}_\varphi(k)\tilde{\CO}_\varphi(-k)\Rangle = i\left(2\Delta-d\right)\frac{\varphi_0}{\tilde{\varphi}_0} &=
	i\left(2\Delta-d\right)\bfrac{\Gamma\big[\Delta-\tfrac{d}{2}\big]}{\Gamma\big[\tfrac{d}{2}-\Delta\big]}\left(\frac{k}{2}\right)^{-2\Delta+d} \\
&\equiv
	-i\left(2\tilde{\Delta}-d\right)\bfrac{\Gamma\big[\tfrac{d}{2}-\tilde{\Delta}\big]}{\Gamma\big[\tilde{\Delta}-\tfrac{d}{2}\big]}\left(\frac{k}{2}\right)^{2\tilde{\Delta}-d}\ ,
\end{align*}
and Fourier transforming we get
\begin{equation}
\Langle\CO_\varphi(x)\CO_\varphi(0)\Rangle = -i\,\bfrac{2\tilde{\Delta}-d}{\pi^\frac{d}{2}}\bfrac{\Gamma\big[\tilde{\Delta}\big]}{\Gamma\big[\tilde{\Delta}-\tfrac{d}{2}\big]}\; \bfrac{1}{|x|^{2\tilde{\Delta}}}\ .
\end{equation}
So, the expressions of the two-point correlator in alternative quantization are identical, both in momentum and position space, to those in standard quantization, provided that we switch $\Delta$ with~$\tilde{\Delta}$. This latter indeed represents the scaling dimension of the dual operator in alternative quantization.

\subsubsection[Holographic renormalization in presence of logarithms]{Holographic renormalization in presence of logarithms: scheme dependence and anomalies}	\label{HoloRenlog}

Let us conclude this crash course on holographic renormalization by briefly considering a case where logarithms arise. Logarithmic terms are a bit more annoying from the technical point of view, but they can be treated in a standard way without any conceptual issue. For the equation of motion~\eqref{eomvarphi}, logarithmic terms appear in the expansion~\eqref{varphiasympt} whenever $2\Delta-d$ is an even integer (zero included). Let us distinguish two different situations: when we are at the BF bound, and so $\Delta=d-\Delta=\tfrac{d}{2}$, and when we are away from it. 

In the former situation, the logarithmic term is the leading, and it is the source in standard quantization, meaning with this that a mass-like counterterm~\eqref{Sct0varphi} is employed to renormalize the boundary action; by means of the corresponding Legendre transformed counterterm, the source can be switched to the term without logarithm, but in both cases an operator of dimension~$\frac{d}{2}$ is described. In the latter situation, instead, the source and response are not affected by logarithmic terms, which on the contrary appear in the sub-leading divergences.

In both situations, the renormalization works in the same way as described above, and counterterms have the same form, except that they may need to be dressed with logarithms where necessary, in order to remove logarithmic divergences. The only crucial consequence of logarithmic counterterms is that they introduce scheme dependent terms in the renormalized action (the coefficient of such terms can be modified by arbitrary finite counterterms).

Let us consider again our real scalar field in AdS, for the case~$2\Delta-d=2$ (other cases are not conceptually different), and so
\begin{equation}
\varphi= r^{\frac{d}{2}-1}\left(\varphi_0 +r^2\ln\!r\,\varphi_1 +\;\ldots\; \right) +r^{\frac{d}{2}+1}\left(\tilde{\varphi}_0 +\;\ldots\; \right)\ .
\end{equation}
With this asymptotic expansion, the on-shell action~\eqref{Sregvarphi} becomes
\begin{equation}	\label{Sregvarphilog}
{}^{\mathrm{(log)}}\!S_{\mathrm{reg}} = \bfrac12\intd[r=\epsilon\!]{d}\,
\left[r^{-2}\left(\bfrac{d}{2}-1\right)\varphi_0 +\ln\!r\;d\,\varphi_1 +d\,\tilde{\varphi}_0 +\varphi_1\right]\varphi_0\ .
\end{equation}
We can notice that, besides the log-divergent term, an additional finite term is present. The equation of motion~\eqref{eomvarphi} relates~$\varphi_1$ to the source, as in~\eqref{varphi1BOXvarphi0}, which though now reads
\begin{equation}	\label{varphi1log}
\varphi_1=-\bfrac{1}{2}\,\Box\varphi_0\ .
\end{equation}
Thus, a counterterm of the form~\eqref{Sct1}, but dressed with an additional logarithmic divergent factor, 
\begin{equation}	\label{Sct1log}
\bfrac{d}{2}\ln(r)\,S_\mathrm{ct}^{(1)} =
		\bfrac{d}{4}\intd{d}\left(\ln\!r\,\varphi_0\Box\varphi_0\right) =
			-\bfrac{d}{2}\intd{d}\,\ln\!r\,\varphi_0\varphi_1\ ,
\end{equation}
is able to remove the logarithmic divergence in the the regularized action~\eqref{Sregvarphilog}. However, we are still allowed to add (finite) counterterms of the form of~$S_\mathrm{ct}^{(1)}$~\eqref{Sct1} (that is, without logarithmic factor). Such counterterms, which are related to the holographic \emph{matter} conformal anomaly~\cite{deHaro:2000vlm,Petkou:1999fv,Mir:2013pca}, would modify the coefficient of the finite term proportional to~$\varphi_0\varphi_1$ in~\eqref{Sregvarphilog}, which can be even removed: such term depends therefore on the chosen renormalization scheme. 

On the other hand, there is no way to touch the term proportional to~$\varphi_0\tilde{\varphi}_0$ by means of finite counterterms: the coefficient of such term is fixed once and for all by the leading counterterm~$S^{(0)}_\mathrm{ct}$~\eqref{Sct0varphi}. Then, we can eventually state again that the final (scheme independent) form of the renormalized action is fully determined by the leading counterterm. We finally have
\begin{equation}	\label{Srenvarphilog}
{}^{\mathrm{(log)}}\!S_{\mathrm{ren}} = \bfrac12\intd{d}\,
\varphi_0\left(\tilde{\varphi}_0 +\ln\!\Lambda\, \Box\varphi_0\right)\ ,
\end{equation}
where $\ln\!\Lambda$ is an arbitrary constant, which we can adapt by choosing a certain renormalization scheme, and whose logarithmic form will be justified in a second. Let us first notice that the expressions for the one-point function~\eqref{1point} gets shifted by the scheme-dependent piece of the renormalized action~\eqref{Srenvarphilog}.

We analyze now the explicit solution, which, for the Fourier transform, can be read off from~\eqref{solvarphi}, replacing the current value for~$\Delta$:
\begin{equation}	\label{solvarphilog}
\varphi_k = C_0\,r^{\frac{d}{2}}\;\sfK[1;\,kr\big]\ ,
\end{equation}
with $C_0$ a constant. The expansions near~$r=0$ gives
\begin{equation}
\varphi_k \sim r^{\frac{d}{2}-1}\;\frac{C_0}{k} 
	+r^{\frac{d}{2}+1}\ln\!r\;C_0\bfrac{k}{2} 
		+r^{\frac{d}{2}+1}\;C_0\bfrac{k}{2}\left(\ln\!\bfrac{k}{2}-\bfrac{1}{2}+\gamma_{EM}\right)\ ,
\end{equation}
where~$\gamma_{EM}$ is the Euler-Mascheroni constant, and from where we read the following expressions for the coefficients:
\begin{align}
\varphi_0 &=
	\frac{C_0}{k}\ ,\\
\tilde{\varphi}_0 &=
	C_0\bfrac{k}{2}\left(\ln\!\bfrac{k}{2}-\bfrac{1}{2}+\gamma_{EM}\right)\ ,\\
\varphi_1 &=
	C_0\bfrac{k}{2}\equiv \bfrac12 k^2\varphi_0\ ,
\end{align}
where in the last line we have pointed out that the analytic expression for~$\varphi_1$ correctly reproduces~\eqref{varphi1log}.

Finally using the formula~\eqref{holocorrfuncts} on the renormalized action~\eqref{Srenvarphilog}, we obtain the two-point function
\begin{equation}
\Langle\CO_\varphi(k)\CO_\varphi(-k)\Rangle=
	-i\bfrac{k^2}{4}\left(\ln\!\bfrac{k^2}{\Lambda^2}-1+2\gamma_{EM}-\ln\!4\right)\ .
\end{equation}
We see here that the meaning of $\Lambda$: it is a scale, introduced by the conformal anomaly, which makes the argument of the logarithm dimensionless.

\bigskip

We conclude this section by admitting that here we have presented holographic renormalization in a very operational way. We do not pretend to be exhaustive about the subject, and we point out that a more formal and general approach exists, for instance those based on Hamiltonian formalism and renormalization group flows~\cite{Kraus:1999di,deBoer:1999tgo,deBoer:2000cz,Martelli:2002sp,Papadimitriou:2004ap,Papadimitriou:2004rz} (see~\cite{Papadimitriou:2016yit} for a pedagogical review).

\clearpage{\pagestyle{empty}\cleardoublepage}
\section[A paradigmatic example]{A paradigmatic example: spontaneous and explicit breaking in a holo\-graph\-ic-dual relativistic CFT}	\label{relativisticsetups}
\label{HoloPGB}

In this section, in order to illustrate the procedure of applying holographic renormalization to derive Ward identities through a simple example, a minimal holographic model is presented, following the analysis performed in~\cite{Argurio:2015wgr}. It describes a quantum field theory where a continuous global symmetry is broken, both spontaneously and explicitly at the same time, by a unique charged operator. 

Thus, by the procedure of holographic renormalization illustrated in Section~\ref{HoloRen}, we retrieve the Ward identity~\eqref{WIsb}, for concomitant spontaneous and explicit breaking. For some specific values of space-time dimensions and dimension of the dual scalar operator, we provide an explicit analytic solution for the two-point scalar correlator, exhibiting the pseudo-Goldstone pole, which correctly reproduce the GMOR relation~\eqref{gmor}.

As introduced in the previous section, eq.~\eqref{currentsource}, a global conserved current in the boundary theory is dual to a local gauge field in the bulk. It is then natural to add a scalar that couples to the gauge field in the bulk, in order to have a scalar operator that is charged under the global symmetry on the boundary. The most basic bulk action that we can write with these two fundamental ingredients\footnote{%
	For the reader familiar with the literature of the holographic superconductor, a disclaimer is here in order. The action~\eqref{SholoPGB} is identical to the one of the very first holographic superconductor~\cite{Gubser:2008px,Hartnoll:2008vx}. However, besides the fact that we will not switch on a background for the vector, since we do not want to break Lorentz invariance having a chemical potential, our purposes are completely different here. In the holographic superconductor the aim is to describe a system at finite temperature and to furnish a dynamical mechanism which generates a critical scale where an order parameter appears and a phase transition occurs. In our case, we want just to mimic an ordinary relativistic field theory at zero temperature, which enjoys a global symmetry, which in turns is broken (spontaneously and/or explicitly) by a scalar operator. So, in a sense, we are always in the broken phase.
} is then
\begin{equation}	\label{SholoPGB}
S=\intd{d+1} \g\left[ -\frac14 F^{mn}F_{mn}-D_m\phi^* D^m\phi -m_\phi^2\phi^*\phi \right]\ ,
\end{equation}
where $F_{mn}=\partial_mA_n-\partial_nA_m$, $D_m\phi=\partial_m\phi-iA_m\phi$, and $g_{mn}$ is the AdS$_{d+1}$ metric~\eqref{AdSmetric}. For our quest of maximal simplicity, we choose an abelian~U$(1)$. Non-abelian symmetries can of course be treated as well, as for instance it is done in reference~\cite{Argurio:2015via}. Here, they would add complications without increasing generality.

The AdS metric is chosen to be a fixed background here and gravity is kept non-dynamical, since we are not going to compute correlation functions which involve the stress-energy tensor. Moreover, we are allowed to neglect the back-reaction of the other fields on the metric because we will carry out a near-boundary analysis, and the back-reaction would emerge at higher order in the near-boundary expansions with respect to the computations that we will perform.

In the introductory section on holographic renormalization, we have pointed out that a vacuum expectation value for the boundary operator corresponds to a fixed profile for the response-mode~\eqref{1point}. On the other hand, a fixed value for the source-mode would correspond to a deformation of the boundary QFT by an operator of dimension $\Delta$ (in standard quantization; $d-\Delta$ in alternative quantization). If the operator is charged under the symmetry, as in our case, due to the coupling to the gauge field, then such deformation breaks the symmetry explicitly~\cite{Balasubramanian:1998de,Klebanov:1999tb,Girardello:1999bd}. So, giving a background profile to our complex scalar will allow to trigger spontaneous and explicit breaking of the boundary global symmetry.

Let us then consider the equation of motion for the complex scalar alone, which are coming from the variation of the action~\eqref{SholoPGB} with respect to~$\phi$, setting~$A_m\equiv0$. It reads
\begin{equation}	\label{eomphiB}
r^{d+1}\partial_r\Big(r^{-d+1}\partial_r\phi\Big) +r^2\Box\phi -m_\phi^2\phi =0\ ,
\end{equation}
which is the same equation as for the real scalar~\eqref{eomvarphi}. Thus the asymptotic solution is
\begin{equation}	\label{phiasympt}
\phi= r^{\frac{d}{2}-\nu}\left(\phiz +\;\ldots\; \right) +r^{\frac{d}{2}+\nu}\left(\tphiz +\;\ldots\; \right)\ , \quad
\text{with }\ 	\nu=\nud\ .
\end{equation}
So, in order to arouse the desired spontaneous and explicit breaking on the boundary, we should give background values to the leading and the sub-leading, namely
\begin{equation}	\label{phiB}
\phi_B= m\,r^{\frac{d}{2}-\nu}+v\,r^{\frac{d}{2}+\nu}\ .
\end{equation}
In standard quantization, the leading piece (source) is triggering explicit breaking, whereas the sub-leading one (vev) is triggering spontaneous breaking. Of course, in alternative quantization source and vev are switched, and so are explicit and spontaneous breaking accordingly.

The Ward identity structure for symmetry breaking in the boundary field theory emerges through the precise holographic renormalization procedure, which therefore constitutes our first task. Let us stay in the window $0<\nu<1$, where no scheme dependence driven by logarithmic terms arise in the holographic renormalization procedure, and in addition it is the window where alternative quantization is possible~\eqref{altquant}. This restriction will not decrease the generality of our conclusions. 

We then apply holographic renormalization to the action~\eqref{SholoPGB}, for fluctuations above the background~\eqref{phiB}, and we divide also the fluctuations of the complex scalar into real and imaginary part,
\begin{equation}
\phi=\bfrac{1}{\sqrt2}\left(\phi_B +\rho + i \pi\right)\ ,
\end{equation}
where $\phi_B$ is assumed to be real for simplicity (and, as we have already remarked in Section~\ref{explicitWI}, on page~\pageref{vrealmreal}, also for consistency). The rescaling pre-factor $\sqrt{2}$ with respect to the generic shape of the scalar profile~\eqref{phiB} is designed to match the field theory derivations of Section~\ref{explicitWI}.

We partially fix the gauge freedom by setting ourselves in the radial gauge~$A_r=0$, leaving unbroken the boundary gauge invariance only. In such situation, the equation of motion for~$A_r$ gives a constraint on the other fields, namely 
\begin{equation}	\label{HPGBeomAr2}
r^2\partial_r \partial_\mu A^\mu -(\phi_B+\rho)\partial_r\pi +\pi\partial_r(\phi_B+\rho) =0\ .
\end{equation}
The equation of motion for~$A_\mu$, instead, reads
\begin{equation}
r^{d-1}\partial_r\big(r^{-d+3}\partial_rA_\mu\big) +r^2\big(\Box A_\mu -\partial_\mu\partial_\nu A^\nu\big)
	+\big(\phi_B+\rho\big)\partial_\mu\pi -\pi\partial_\mu\rho -(\phi_B+\rho)^2A_\mu =0\ ,	\label{eomAmu}
\end{equation}
from which we can extract the asymptotic behavior
\begin{equation}
A^\mu = a^\mu_0 +r^2a^\mu_1 \;\ldots\; +r^{d-2}\left(\tilde{a}^\mu_0+\;\ldots\;\right)\ .
\end{equation}

We can notice that for~$d=2$ the leading and the sub-leading are of the same order (as for the scalar at the BF bound), and so a logarithmic term appears. This case has been considered in~\cite{Argurio:2016xih}, where the issues related to Goldstone theorem in two-dimensions are discussed. For~$d=4$, a logarithm occurs as well, since $a^\mu_1$ and~$\tilde{a}^\mu_0$ are of same order, but this is less problematic since the source~$a^\mu_0$ is not affected\footnote{%
	For an example of holographic renormalization of the gauge field in~$d=4$, we point the reader to~\cite{Argurio:2014rja}, where it is treated in a similar context.
}. However, in this section, we will set~$d=3$, so that we do not have to renormalize the vector, and we have no scheme dependence for either the vector and the scalar. Furthermore, we remark that for our chosen window of values for~$\nu$ ($0<\nu<1$), the scalar background intervenes in the equation of motion for~$A_\mu$~\eqref{eomAmu} at the order~$d-2\nu>d-2$, thus not affecting the sub-leading of~$A_\mu$. Otherwise, we could not have avoided considering a background for the vector as well.

So, we stay at $d=3$, and in addition, since we are interested in two-point functions at most, we will consider the renormalized action up to quadratic order in the fluctuations. In such case, we can consider the linearized equations of motion for the fluctuated field over~$\phi_B$, which read
\begin{align}
&	
r^4\partial_r\big(r^{-2}\partial_r\rho\big) +r^2\big(\Box -m_\phi^2\big)\rho =0 \ ,	\vphantom{\Big|}\label{HPGBeomrho1}\\
&	
r^4\partial_r\big(r^{-2}\partial_r\pi\big) +r^2\big(\Box -m_\phi^2\big)\pi -r^2\phi_B\partial_\mu A^\mu =0 \ ,	\vphantom{\Big|}\label{HPGBeompi1}\\
& 
r^2\partial_r^2A_\mu +r^2\big(\Box A_\mu -\partial_\mu\partial_\nu A^\nu\big) +\phi_B\partial_\mu\pi -\phi_B^2A_\mu =0\ , \vphantom{\Big|}\label{HPGBeomAmu1}\\
&
r^2\partial_r \partial_\mu A^\mu -\phi_B\partial_r\pi +\pi\partial_r\phi_B =0\label{HPGBeomAr1}\ .
\end{align}
Then, the on-shell action at quadratic order in the fluctuations reduces to the following boundary term
\begin{equation}
S_\mathrm{reg}=\intd[r=\epsilon\!]{3} \left[ \bfrac{1}{r^2}(\partial_r\phi_B)\rho +\bfrac12 A^\mu \partial_r A_\mu +\bfrac{1}{2r^2}\left(\rho\partial_r\rho+\pi\partial_r\pi \right)\right]\ .\label{HPGBSreg}
\end{equation}
At this point, we note that the quadratic terms are exactly the same that would arise in a configuration with vanishing backgrounds. The presence of a non-trivial background must then show up when expanding the fluctuations near the boundary as powers of $r$. There is however one more substitution that we can make, that makes the dependence on the background manifest even before expanding the fluctuations. We can indeed use the equation of motion coming from the variation with respect to~$A_r$~\eqref{HPGBeomAr1}, which at linear order in the fluctuations rewrites
\begin{equation} \label{HPGBeomAr}
r^2\partial_r\Box L - \phi_B\partial_r\pi+\partial_r\phi_B\pi=0\ ,
\end{equation}
where we have also introduced the splitting of the gauge field in its irreducible components,
\begin{equation}	\label{gaugesplit}
A_\mu=T_\mu+\partial_\mu L\ , \qquad	\text{with }\ \partial_\mu T^\mu=0\ .
\end{equation}
Noting that the second term of \eqref{HPGBSreg} has a longitudinal part that can be rewritten, 
after integration by parts, as $\frac12 L \partial_r \Box L$, then the regularized action for the longitudinal part and the scalars becomes
\begin{equation} \label{HPGBSreg2} 
S_\mathrm{reg}=\bfrac12 \intd{3} \Big[ 
	T^\mu\partial_rT_\mu\ +\bfrac{2}{r^2}\,\rho\partial_r\phi_B +\bfrac{1}{r^2}L\left(\pi\partial_r\phi_B-\phi_B\partial_r\pi\right) +\bfrac{1}{r^2}\left(\rho\partial_r\rho+\pi\partial_r\pi \right)\Big]\ .
\end{equation}

Using the splitting~\eqref{gaugesplit}, we now rewrite the equation of motion for the imaginary part of the scalar~\eqref{HPGBeompi1},
\begin{equation}
r^4\partial_r\big(r^{-2}\partial_r\pi\big) +r^2\big(\Box -m_\phi^2\big)\pi -r^2\phi_B\Box L =0 \ ,	\label{HPGBeompi}\\
\end{equation}
and the equation of motion for the vector field~\eqref{HPGBeomAmu1},
\begin{align}
& 
	r^2\partial_r^2L +\phi_B\pi -\phi_B^2L =0\ , \vphantom{\Big|}\label{HPGBeomL}\\
&
	r^2\partial_r^2T_\mu +r^2\Box T_\mu -\phi_B^2T_\mu =0\ , \vphantom{\Big|}\label{HPGBeomT}
\end{align}
where we see that the equation for the transverse part of the vector decouples, exactly as for the real part of the scalar~\eqref{HPGBeomrho1}. Then, from the equations of motion we can derive the following asymptotic expansions for the fluctuated fields:
\begin{equation}\label{nbe}
\begin{aligned}
\rho &= r^{\frac{3}{2}-\nu}\,\rhoz +r^{\frac{3}{2}+\nu}\,\trhoz +\;\dots\ , \quad&
	T^\mu &= t_0^\mu +r\,\tilde{t}_0^{\,\mu} +\;\dots\ ,	\quad\\
\pi &= r^{\frac{3}{2}-\nu}\,\piz +r^{\frac{3}{2}+\nu}\,\tpiz +\;\dots\ , \quad&
	L &= l_0+r\,\tilde{l}_0 +\;\dots\ . 	\quad
\end{aligned}
\end{equation}
The regularized action~\eqref{HPGBSreg2} then becomes
\begin{align}
S_\mathrm{reg}= \bfrac{1}{2}\intd{3} \Big[&
	t_0\cdot\tilde{t}_0 +(3-2\nu)\,m\big(r^{-2\nu}\rho_0 +\trhoz\big) +(3+2\nu)\,v\rho_0 \;+		\label{HPGBSreg3}\\
&
	+\Big(\bfrac32-\nu\Big)r^{-2\nu}\left(\rho_0^2+\pi_0^2\right) +3\left(\rho_0\trhoz+\pi_0\tpiz\right) +2\nu\,l_0\big(v\pi_0-m\tpiz\big)
\Big]\ .\nonumber
\end{align}
The counterterm needed to cancel the divergences in standard quantization is analogous to the one for the real scalar~\eqref{Sct0varphi}, and reads
\begin{equation}
S_\mathrm{ct}=-\Big(\bfrac32-\nu\Big)\intd{3}\, \gh\; \phi^*\phi = 
	-\bfrac12\Big(\bfrac32-\nu\Big)\intd{3}\, r^{-3}\left( 2\phi_B\rho +\rho^2+\pi^2\right)  \ .\label{sct}
\end{equation}
Note that we  neglect the constant term, as it would only be relevant with dynamical gravity.
After adding the counterterm~\eqref{sct} to the regularized action~\eqref{HPGBSreg3}, we obtain the renormalized action
\begin{equation}
S_\mathrm{ren}= 2\nu\intd{3} \left[ v\rho_0 +\bfrac12\trhoz\rho_0
+\bfrac12\tpiz\big(\pi_0-ml_0\big) +\bfrac12vl_0\pi_0\right]\ ,\label{HPGBSren0}
\end{equation}
where we have dropped the term for the transverse part of the vector, since it is completely decoupled from the scalar sector. 

We can then notice that there are two kinds of terms in the quadratic renormalized action: those which are bilinears of a source and a response of the fluctuations, and those which involve only sources. The latter are all proportional to the non-trivial scalar background that we have introduced, thus they would not be there for trivial profile~$v=0=m$. In practice, the background profiles make the scalar and vector sectors `talk' to each others. 

However, terms of the second kind are also hidden into terms of the first kind, because of gauge invariance. Indeed, gauge transformations that preserve our gauge choice~$A_r=0$ yield
\begin{equation}
\delta L=\alpha, \qquad \delta \phi=i\alpha \phi\ .
\end{equation}
The first transformation above tells that $\alpha$ should be considered of the same order as the 
fluctuations $L$ and $\rho, \pi$. It then implies that the gauge variations of $\rho,\pi$ have actually terms of first and second order 
\begin{equation}
\delta\rho = -\alpha\pi \ , \qquad \delta\pi =  \alpha\phi_B + \alpha \rho\ .
\end{equation}
On the coefficients of the near-boundary expansions \eqref{nbe}, the transformations read
\begin{equation}	\label{HPGBgaugetransfs}
\begin{aligned}
\delta l_0 &= \alpha\ , &\quad
	\delta\rho_0 &= -\alpha\pi_0\ , &\quad
		\delta\pi_0 &= \alpha m + \alpha\rho_0\ , \\
\delta\tilde{l}_0 &= 0\ ,	&\quad
	\delta\trhoz &= -\alpha\tpiz\ ,	&\quad
		\delta\tpiz &= \alpha v +\alpha\trhoz\ .
\end{aligned}
\end{equation}
With the transformations given above, one can check that all the actions~$S_\mathrm{reg}$, $S_\mathrm{ct}$, and~$S_\mathrm{ren}$ are gauge invariant. 
We note that gauge invariance requires the cancellation between the variations of 
the linear and quadratic parts of the actions, and we have of course neglected 
orders higher than quadratic%
	\footnote{It is possible also to parametrize the complex scalar in terms of its modulus and phase as in~\cite{Argurio:2014rja}; the latter parametrization, being well-adapted to gauge transformations (which consist in shifts of the phase), features manifest gauge invariance without mixing among different orders in the fluctuations. However, this brings disadvantages in the renormalization procedure: indeed, given that the phase has to be non-dimensional, the would-be Goldstone boson mixes non-trivial with the scalar background~$\phi_B$.}
(\emph{i.e.}~in the variations of the quadratic part of 
the action, only the terms of first order in the variations of $\rho,\pi$ are considered).

We want to the derive the holographic correlators, assuming that the terms coupling the sources to the operators are 
\begin{equation}
\intd{3} \left(\rho_0 \ReO_\phi + \pi_0 \ImO_\phi -l_0 \partial_\mu J^\mu\right)\ ,
\end{equation}
where the last term comes from integration by parts. We also assume the usual holographic prescription in its Wick-rotated, Lorentzian version~\eqref{holocorrfuncts}. In this way, from the renormalized action~\eqref{HPGBSren0} we immediately have that~$\ReO_\phi$ has a non-zero vev, namely
\begin{equation}
\Langle \ReO_\phi\Rangle = \frac{\delta iS_\mathrm{ren}}{\delta i\rho_0\;} = v\ .
\end{equation}
From~\eqref{HPGBSren0}, it is also manifest that the $\rho$-sector decouples from the $L,\pi$-sector. We will focus on this latter, since it is the sector where we expect the (pseudo) Goldstone boson to appear.

In order to solve for $\tpiz$ in terms of the sources $\pi_0$ and $l_0$, one should assure that the deep bulk (IR) boundary conditions preserve gauge invariance, hence they have to be imposed on gauge invariant combinations. At linear order, the gauge invariant combinations are $\pi_0-ml_0$ and $\tpiz-vl_0$. As a consequence, one can express the sub-leading mode of $\pi$ in terms of the sources as
\begin{equation}	\label{nonlocf}
\tpiz= v l_0 +f(\Box)(\pi_0-m l_0)\ .
\end{equation}
The  renormalized action for this sector can be rewritten accordingly
\begin{equation}
S_\mathrm{ren}=-\intd{3} \, \Big[ -\bfrac12(\pi_0-ml_0)f(\Box)(\pi_0-ml_0)
-vl_0\pi_0+\bfrac12mv l_0l_0 \,\Big] 	\ .\label{sren}
\end{equation}
We observe that we have a term that is linear in~$m$, which encodes the operator identities that are present when the symmetry is explicitly broken. Then we have a term linear in~$v$, which embodies the Ward identities when the symmetry is spontaneously broken. Eventually we have a term which is linear both in $m$ and $v$ and is necessary in order to recover the proper Ward identities in the case of concomitant spontaneous and explicit breaking.

Indeed, again using the prescription~\eqref{holocorrfuncts} for deriving two-point functions, we obtain the following relations among correlators of the longitudinal sector:
\begin{align}
\Langle \ImO_\phi\ImO_\phi\Rangle &=  -\bfrac{\delta^2 iS_\mathrm{ren}}{\delta\pi_0\delta\pi_0}=-if(\Box)\ ,\label{corrpi0pi0}\\
\Langle\partial_\mu J^\mu\ImO_\phi\Rangle &=  \bfrac{\delta^2 iS_\mathrm{ren}}{\delta l_0\delta\pi_0}=-imf(\Box)+iv\ , \label{WIJImO}\\
\Langle\partial_\mu J^\mu\partial_\nu J^\nu\Rangle &=  -\bfrac{\delta^2 iS_\mathrm{ren}}{\delta l_0\delta l_0}=-im^2f(\Box)+imv\  .
\end{align}
These are exactly the equations \eqref{oocorr}--\eqref{jjcorr}, obtained in Section~\ref{explicitWI} from QFT arguments, where they have been used to derive the GMOR relation. 

In the original paper~\cite{Argurio:2015wgr}, the non-trivial function~$f(\Box)$ was also computed analytically for some specific values of the dimension of the dual operator, by imposing boundary conditions in the bulk and solving the equation of motion. The consequent analytic control on the expression of the scalar two-point correlator allowed for smoothly moving from the purely spontaneous case (which exhibit the massless Goldstone pole) to the purely explicit one, and in particular to extract an explicit expression of the GMOR relation for the pseudo-Goldstone pole, in the case where~$m\!\ll\!\sqrt{v}$.

\clearpage{\pagestyle{empty}\cleardoublepage}
\section*{Summary and related examples}
\markboth{Summary and related examples}{}
\phantomsection
\addcontentsline{toc}{section}{Summary and related examples}

The main focus of these notes was to illustrate in a pedagogical and practical way the derivation of Ward identities encoding symmetry breaking in holography. A crucial step was covered by the procedure of holographic renormalization.
\medskip

In Section~1, as preliminary material, we reviewed, from a field theory perspective, various features of the physics of symmetry breaking, both in relativistic and in non-relativistic contexts. 

Then, in Section~2, we briefly introduced the holographic correspondence and the prescription for computing field theory correlation functions, and we presented the procedure of holographic renormalization through a paradigmatic example, discussing some peculiarities such as alternative quantization and scheme dependence.

In the third section, we presented a simple and prototypic relativistic bulk toy-model for an abelian U(1) symmetry breaking, where the Ward identity structure and symmetry breaking pattern were neatly realized. The precise relations among correlators are dictated by the field theory arguments of Section~\ref{explicitWI}, which pinpoint the Ward identity structure independently of the strength of the coupling, and are realized in the AdS/CFT model thanks to holographic renormalization. The holographic derivation relies just on an asymptotic near-boundary analysis, therefore, as far as Ward identities are concerned, only UV knowledge is necessary, and we could indeed perform the analysis without actually solving the model. 

Thus holography, already in such a basic realization, is able to reproduce general quantum field theoretical expectations, and, although only for specific space-time dimensions and scaling dimension of the dual scalar operator, allows also for explicit quantitative computations~\cite{Argurio:2015wgr}.
\medskip

Let us conclude by outlining some further analogous examples that have been discussed so far in the literature. The lower dimensional case of two boundary dimensions was discussed in~\cite{Argurio:2016xih}, where the fact that in the strict large~$N$ limit, as we pointed out in Section~\ref{ColemanTh}, spontaneous symmetry breaking can occur in two dimensions has been recovered in holography. Indeed, considering the AdS$_3$/CFT$_2$ version of the model of Section~\ref{HoloPGB},the same Ward identities are retrieved as they appear in higher dimensions. Nevertheless, the way to get this result involves subtleties and peculiarities which are specific to two dimensions, and can be regarded as a premonition of the fact that spontaneous breaking is ruled out as soon as one moves away from strict infinite $N$. The most crucial subtlety is that the gauge field has to be renormalized in a sort of `alternative quantization', in order to have it properly sourcing a conserved current and yielding the correct Ward identities for the breaking of a global symmetry on the boundary.

However, if the existence of Goldstone bosons in two dimensions is under threat, that is absolutely not the case for pseudo-Goldstone bosons. Indeed, in $1\!+\!1$ dimensions, the large quantum fluctuations of the phase prevent the selection of a specific ground state out of the continuum of possible vacua; however, if we added an arbitrarily small (but finite) explicit breaking, this would select a particular ground state, and act as a regulator for the infra-red divergence, making such vacuum stable under quantum fluctuations. Hence, for explicit breaking parametrically smaller than the spontaneous one, we expect (even at finite $N$) a mode that is hierarchically lighter than the rest of the spectrum, and whose mass undergoes the usual GMOR relation. So, there is no obstruction for pseudo-Goldstone bosons in two dimensions, and reference~\cite{Argurio:2016xih} provides a holographic model for them.

Then, one can take into account non-relativistic setups. In reference~\cite{Argurio:2015via}, a non-abelian version (with a global U(2) symmetry) of the relativistic model of Section~\ref{HoloPGB} is considered, and Lorentz invariance is broken by giving a background profile to the temporal component of the gauge field. Performing the holographic renormalization of this setup, the Ward identities associated to the broken symmetries are reproduced in a non-relativistic framework. So the system must have Goldstone excitations associated to the three broken generators. However, because of broken Lorentz symmetry, not only the scalar operator acquired a symmetry breaking vev's, but also the temporal components of one of the three conserved currents. In such situation so-called type B Goldstone bosons arise: the commutator of two broken charges has a non-vanishing vev, and for these two conjugate broken generators only one massless excitation appears, which in turn typically possesses quadratic dispersion relation. The third broken generator is not involved in commutators with non-trivial VEVs, and hence gives rise to a type~A Goldstone boson.

From the field theoretical analysis of Section~\ref{nonrelGoldstoneTh}, we expect the type~B GB to have quadratic dispersion relation and the type~A GB to have linear dispersion relation, but with a velocity smaller than the speed of light~$c$. Moreover, we expect the type~B Goldstone boson to be accompanied by an almost Goldstone boson, \emph{i.e.}~a light mode whose mass is related to the coefficient of the quadratic dispersion relation of its partner \cite{Kapustin:2012cr}. Unfortunately, due to littler symmetry and a much more involved system of equations of motion, contrary to the model of~\cite{Argurio:2015wgr}, no analytic expressions containing such a spectrum were found in~\cite{Argurio:2015via}.

Furthermore, in a more recent paper~\cite{Argurio:2017xxx} conserved currents and charged scalar operators in Lifshitz invariant space-times are studied. Again, performing in detail the holographic renormalization for the most general bulk action of a gauge vector field coupled to a charged scalar on a fixed Lifshitz background, where the scalar field is given a profile in order to trigger symmetry breaking on the boundary, the proper non-relativistic realization of Ward identities for symmetry breaking are obtained.
\smallskip

The study of these various holographic setups yields interesting insights on the physics of symmetry breaking in quantum field theory, especially in the less well-defined non-relativistic framework. At the same time, it gives the occasion to explore the holographic correspondence in a variety of canonical and less canonical examples. The implementation of holographic renormalization to those various examples may encounter some specificities and subtleties. We hope that these notes served as a good starting point for the reader to handle them.

\subsection*{Aknowledgements}
\noindent
Let me thank first Riccardo Argurio, for his continuous support as advisor along all my PhD studies. Then, together with him, I thank as well the other collaborators with whom and through whom I have learned what I know about the topics discussed here: Andrea Mezzalira, Daniele Musso, Daniel Naegels. I am grateful to St\'ephane Detournay, Nabil Iqbal and Marika Taylor, for their precious feedbacks as readers of my thesis. I finally thank Tom\'a\v{s} Brauner for some clarifying exchanges about Goldstone theorem in lack of Lorentz invariance.\\ 
This work was supported by IISN-Belgium (convention 4.4503.15).
\bigskip
\bigskip

\pagestyle{fancy}
\fancyhead{}
\rhead[\fancyplain{}{\scshape\leftmark}]{\fancyplain{}{\thepage}}
\lhead[\fancyplain{}{\thepage}]{\fancyplain{}{\scshape\rightmark}}
\bibliography{bibliography}

\providecommand{\href}[2]{#2}\begingroup\raggedright\begin{thebibliography}{10}

\bibitem{Nambu:1960tm}
Y.~Nambu, ``{{Quasiparticles and Gauge Invariance in the Theory of
  Superconductivity}}'',
\href{http://dx.doi.org/10.1103/PhysRev.117.648}{{\em Phys. Rev.} {\bfseries
  117} (1960) 648--663}.

\bibitem{Goldstone:1961eq}
J.~Goldstone, ``{{Field Theories with Superconductor Solutions}}'',
\href{http://dx.doi.org/10.1007/BF02812722}{{\em Nuovo Cim.} {\bfseries 19}
  (1961) 154--164}.

\bibitem{Goldstone:1962es}
J.~Goldstone, A.~Salam, and S.~Weinberg, ``{{Broken Symmetries}}'',
\href{http://dx.doi.org/10.1103/PhysRev.127.965}{{\em Phys. Rev.} {\bfseries
  127} (1962) 965--970}.

\bibitem{Klein:1964ix}
A.~Klein and B.~W. Lee, ``{{Does Spontaneous Breakdown of Symmetry Imply
  Zero-Mass Particles?}}'',
\href{http://dx.doi.org/10.1103/PhysRevLett.12.266}{{\em Phys. Rev. Lett.}
  {\bfseries 12} (1964) 266--268}.

\bibitem{Lange:1965zz}
R.~V. Lange, ``{{Goldstone Theorem in Nonrelativistic Theories}}'',
\href{http://dx.doi.org/10.1103/PhysRevLett.14.3}{{\em Phys. Rev. Lett.}
  {\bfseries 14} (1965) 3--6}.

\bibitem{Lange:1966zz}
R.~V. Lange, ``{{Nonrelativistic Theorem Analogous to the Goldstone
  Theorem}}'',
\href{http://dx.doi.org/10.1103/PhysRev.146.301}{{\em Phys. Rev.} {\bfseries
  146} (1966) 301--303}.

\bibitem{Guralnik:1967zz}
G.~S. Guralnik, C.~R. Hagen, and T.~W.~B. Kibble, ``{{Broken symmetries and the
  Goldstone theorem}}'', {\em Adv. Part. Phys.} {\bfseries 2} (1968) 567--708.
\url{http://www.physics.princeton.edu/~mcdonald/examples/EP/guralnik_ap_2_567_67.pdf}.

\bibitem{Nielsen:1975hm}
H.~B. Nielsen and S.~Chadha, ``{{On How to Count Goldstone Bosons}}'',
\href{http://dx.doi.org/10.1016/0550-3213(76)90025-0}{{\em Nucl. Phys.}
  {\bfseries B105} (1976) 445--453}.

\bibitem{Brauner:2007uw}
T.~Brauner, ``{{Goldstone bosons in presence of charge density}}'',
  \href{http://dx.doi.org/10.1103/PhysRevD.75.105014}{{\em Phys. Rev.}
  {\bfseries D75} (2007) 105014},
\href{http://arxiv.org/abs/hep-ph/0701110}{{\ttfamily arXiv:hep-ph/0701110
  [hep-ph]}}.

\bibitem{Watanabe:2011ec}
H.~Watanabe and T.~Brauner, ``{On the number of {Nambu-Goldstone} bosons and
  its relation to charge densities}'',
  \href{http://dx.doi.org/10.1103/PhysRevD.84.125013}{{\em Phys. Rev.}
  {\bfseries D84} (2011) 125013},
\href{http://arxiv.org/abs/1109.6327}{{\ttfamily arXiv:1109.6327 [hep-ph]}}.

\bibitem{Watanabe:2012hr}
H.~Watanabe and H.~Murayama, ``{Unified Description of {Nambu-Goldstone} Bosons
  without {Lorentz} Invariance}'',
  \href{http://dx.doi.org/10.1103/PhysRevLett.108.251602}{{\em Phys. Rev.
  Lett.} {\bfseries 108} (2012) 251602},
\href{http://arxiv.org/abs/1203.0609}{{\ttfamily arXiv:1203.0609 [hep-th]}}.

\bibitem{Kapustin:2012cr}
A.~Kapustin, ``{{Remarks on nonrelativistic Goldstone bosons}}'',
\href{http://arxiv.org/abs/1207.0457}{{\ttfamily arXiv:1207.0457 [hep-ph]}}.

\bibitem{Maldacena:1997re}
J.~M. Maldacena, ``{{The Large N limit of superconformal field theories and
  supergravity}}'', \href{http://dx.doi.org/10.1023/A:1026654312961}{{\em Int.
  J. Theor. Phys.} {\bfseries 38} (1999) 1113--1133},
  \href{http://arxiv.org/abs/hep-th/9711200}{{\ttfamily arXiv:hep-th/9711200
  [hep-th]}}.
[Adv. Theor. Math. Phys.2,231(1998)].

\bibitem{Son:2008ye}
D.~T. Son, ``{{Toward an AdS/cold atoms correspondence: A Geometric realization
  of the Schrodinger symmetry}}'',
  \href{http://dx.doi.org/10.1103/PhysRevD.78.046003}{{\em Phys. Rev.}
  {\bfseries D78} (2008) 046003},
\href{http://arxiv.org/abs/0804.3972}{{\ttfamily arXiv:0804.3972 [hep-th]}}.

\bibitem{Balasubramanian:2008dm}
K.~Balasubramanian and J.~McGreevy, ``{{Gravity duals for non-relativistic
  CFTs}}'', \href{http://dx.doi.org/10.1103/PhysRevLett.101.061601}{{\em Phys.
  Rev. Lett.} {\bfseries 101} (2008) 061601},
\href{http://arxiv.org/abs/0804.4053}{{\ttfamily arXiv:0804.4053 [hep-th]}}.

\bibitem{Kachru:2008yh}
S.~Kachru, X.~Liu, and M.~Mulligan, ``{{Gravity duals of {Lifshitz-like} fixed
  points}}'', \href{http://dx.doi.org/10.1103/PhysRevD.78.106005}{{\em Phys.
  Rev.} {\bfseries D78} (2008) 106005},
\href{http://arxiv.org/abs/0808.1725}{{\ttfamily arXiv:0808.1725 [hep-th]}}.

\bibitem{Taylor:2008tg}
M.~Taylor, ``{{Non-relativistic holography}}'',
\href{http://arxiv.org/abs/0812.0530}{{\ttfamily arXiv:0812.0530 [hep-th]}}.

\bibitem{Guica:2010sw}
M.~Guica, K.~Skenderis, M.~Taylor, and B.~C. van Rees, ``{Holography for
  {Schr{\"o}dinger} backgrounds}'',
  \href{http://dx.doi.org/10.1007/JHEP02(2011)056}{{\em JHEP} {\bfseries 02}
  (2011) 056},
\href{http://arxiv.org/abs/1008.1991}{{\ttfamily arXiv:1008.1991 [hep-th]}}.

\bibitem{Taylor:2015glc}
M.~Taylor, ``{{Lifshitz holography}}'', {\em Class. Quant. Grav.} {\bfseries
  33} no.~3, (2016) 033001,
\href{http://arxiv.org/abs/1512.03554}{{\ttfamily arXiv:1512.03554 [hep-th]}}.

\bibitem{deHaro:2000vlm}
S.~de~Haro, S.~N. Solodukhin, and K.~Skenderis, ``{{Holographic reconstruction
  of space-time and renormalization in the AdS / CFT correspondence}}'',
  \href{http://dx.doi.org/10.1007/s002200100381}{{\em Commun. Math. Phys.}
  {\bfseries 217} (2001) 595--622},
\href{http://arxiv.org/abs/hep-th/0002230}{{\ttfamily arXiv:hep-th/0002230
  [hep-th]}}.

\bibitem{Bianchi:2001de}
M.~Bianchi, D.~Z. Freedman, and K.~Skenderis, ``{How to go with an {RG}
  flow}'', \href{http://dx.doi.org/10.1088/1126-6708/2001/08/041}{{\em JHEP}
  {\bfseries 08} (2001) 041},
\href{http://arxiv.org/abs/hep-th/0105276}{{\ttfamily arXiv:hep-th/0105276
  [hep-th]}}.

\bibitem{Bianchi:2001kw}
M.~Bianchi, D.~Z. Freedman, and K.~Skenderis, ``{{Holographic
  renormalization}}'',
  \href{http://dx.doi.org/10.1016/S0550-3213(02)00179-7}{{\em Nucl. Phys.}
  {\bfseries B631} (2002) 159--194},
\href{http://arxiv.org/abs/hep-th/0112119}{{\ttfamily arXiv:hep-th/0112119
  [hep-th]}}.

\bibitem{Weinberg:1996kr}
S.~Weinberg, {\em {The quantum theory of fields. Vol. 2: Modern applications}}.
\newblock Cambridge University Press,
2013.
\newblock

\bibitem{Coleman:1973ci}
S.~R. Coleman, ``{{There are no Goldstone bosons in two-dimensions}}'',
\href{http://dx.doi.org/10.1007/BF01646487}{{\em Commun. Math. Phys.}
  {\bfseries 31} (1973) 259--264}.

\bibitem{Ma:1974tp}
S.-K. Ma and R.~Rajaraman, ``{{Comments on the Absence of Spontaneous Symmetry
  Breaking in Low Dimensions}}'',
\href{http://dx.doi.org/10.1103/PhysRevD.11.1701}{{\em Phys. Rev.} {\bfseries
  D11} (1975) 1701}.

\bibitem{Witten:1978qu}
E.~Witten, ``{Chiral Symmetry, the {$1/n$} Expansion, and the {SU$(N)$
  Thirring} Model}'',
\href{http://dx.doi.org/10.1016/0550-3213(78)90416-9}{{\em Nucl. Phys.}
  {\bfseries B145} (1978) 110--118}.

\bibitem{GellMann:1968rz}
M.~Gell-Mann, R.~J. Oakes, and B.~Renner, ``{Behavior of current divergences
  under {SU(3) x SU(3)}}'',
\href{http://dx.doi.org/10.1103/PhysRev.175.2195}{{\em Phys. Rev.} {\bfseries
  175} (1968) 2195--2199}.

\bibitem{Argurio:2015wgr}
R.~Argurio, A.~Marzolla, A.~Mezzalira, and D.~Musso, ``{{Analytic
  pseudo-Goldstone bosons}}'',
  \href{http://dx.doi.org/10.1007/JHEP03(2016)012}{{\em JHEP} {\bfseries 03}
  (2016) 012},
\href{http://arxiv.org/abs/1512.03750}{{\ttfamily arXiv:1512.03750 [hep-th]}}.

\bibitem{Soldati1}
R.~Soldati, ``{Introduction to Quantum Field Theory (A Primer for a Basic
  Education)}''.
\newblock \url{http://www.robertosoldati.com/archivio/news/107/Campi1.pdf}.

\bibitem{Brauner:2010wm}
T.~Brauner, ``{{Spontaneous Symmetry Breaking and Nambu-Goldstone Bosons in
  Quantum Many-Body Systems}}'',
  \href{http://dx.doi.org/10.3390/sym2020609}{{\em Symmetry} {\bfseries 2}
  (2010) 609--657},
\href{http://arxiv.org/abs/1001.5212}{{\ttfamily arXiv:1001.5212 [hep-th]}}.

\bibitem{Witten:1998qj}
E.~Witten, ``{{Anti-de Sitter space and holography}}'', {\em Adv. Theor. Math.
  Phys.} {\bfseries 2} (1998) 253--291,
\href{http://arxiv.org/abs/hep-th/9802150}{{\ttfamily arXiv:hep-th/9802150
  [hep-th]}}.

\bibitem{Polchinski:2010hw}
J.~Polchinski,
  \href{http://dx.doi.org/10.1142/9789814350525_0001}{``{{Introduction to
  Gauge/Gravity Duality}}'',} in {\em {Proceedings, Theoretical Advanced Study
  Institute in Elementary Particle Physics (TASI 2010). String Theory and Its
  Applications: From meV to the Planck Scale: Boulder, Colorado, USA, June
  1-25, 2010}}, pp.~3--46.
\newblock 2010.
\newblock
\href{http://arxiv.org/abs/1010.6134}{{\ttfamily arXiv:1010.6134 [hep-th]}}.
\newblock

\bibitem{Weinberg:1995mt}
S.~Weinberg, {\em {The Quantum theory of fields. Vol. 1: Foundations}}.
\newblock Cambridge University Press,
2005.
\newblock

\bibitem{Low:2001bw}
I.~Low and A.~V. Manohar, ``{{Spontaneously broken space-time symmetries and
  Goldstone's theorem}}'',
  \href{http://dx.doi.org/10.1103/PhysRevLett.88.101602}{{\em Phys. Rev. Lett.}
  {\bfseries 88} (2002) 101602},
\href{http://arxiv.org/abs/hep-th/0110285}{{\ttfamily arXiv:hep-th/0110285
  [hep-th]}}.

\bibitem{Watanabe:2013iia}
H.~Watanabe and H.~Murayama, ``{{Redundancies in Nambu-Goldstone Bosons}}'',
  \href{http://dx.doi.org/10.1103/PhysRevLett.110.181601}{{\em Phys. Rev.
  Lett.} {\bfseries 110} no.~18, (2013) 181601},
\href{http://arxiv.org/abs/1302.4800}{{\ttfamily arXiv:1302.4800
  [cond-mat.other]}}.

\bibitem{Hayata:2013vfa}
T.~Hayata and Y.~Hidaka, ``{{Broken spacetime symmetries and elastic
  variables}}'', \href{http://dx.doi.org/10.1016/j.physletb.2014.06.039}{{\em
  Phys. Lett.} {\bfseries B735} (2014) 195--199},
\href{http://arxiv.org/abs/1312.0008}{{\ttfamily arXiv:1312.0008 [hep-th]}}.

\bibitem{Brauner:2014aha}
T.~Brauner and H.~Watanabe, ``{{Spontaneous breaking of spacetime symmetries
  and the inverse Higgs effect}}'',
  \href{http://dx.doi.org/10.1103/PhysRevD.89.085004}{{\em Phys. Rev.}
  {\bfseries D89} no.~8, (2014) 085004},
\href{http://arxiv.org/abs/1401.5596}{{\ttfamily arXiv:1401.5596 [hep-ph]}}.

\bibitem{Thirring:1958in}
W.~E. Thirring, ``{{A Soluble relativistic field theory?}}'',
\href{http://dx.doi.org/10.1016/0003-4916(58)90015-0}{{\em Annals Phys.}
  {\bfseries 3} (1958) 91--112}.

\bibitem{Argurio:2016xih}
R.~Argurio, G.~Giribet, A.~Marzolla, D.~Naegels, and J.~A. Sierra-Garcia,
  ``{{Holographic Ward identities for symmetry breaking in two dimensions}}'',
  \href{http://dx.doi.org/10.1007/JHEP04(2017)007}{{\em JHEP} {\bfseries 04}
  (2017) 007},
\href{http://arxiv.org/abs/1612.00771}{{\ttfamily arXiv:1612.00771 [hep-th]}}.

\bibitem{Gross:1974jv}
D.~J. Gross and A.~Neveu, ``{{Dynamical Symmetry Breaking in Asymptotically
  Free Field Theories}}'',
\href{http://dx.doi.org/10.1103/PhysRevD.10.3235}{{\em Phys. Rev.} {\bfseries
  D10} (1974) 3235}.

\bibitem{Gasser:1982ap}
J.~Gasser and H.~Leutwyler, ``{{Quark Masses}}'',
\href{http://dx.doi.org/10.1016/0370-1573(82)90035-7}{{\em Phys. Rept.}
  {\bfseries 87} (1982) 77--169}.

\bibitem{Giusti:1998wy}
L.~Giusti, F.~Rapuano, M.~Talevi, and A.~Vladikas, ``{{The {QCD} chiral
  condensate from the lattice}}'',
  \href{http://dx.doi.org/10.1016/S0550-3213(98)00659-2}{{\em Nucl. Phys.}
  {\bfseries B538} (1999) 249--277},
\href{http://arxiv.org/abs/hep-lat/9807014}{{\ttfamily arXiv:hep-lat/9807014
  [hep-lat]}}.

\bibitem{Evans:2004ia}
N.~J. Evans and J.~P. Shock, ``{{Chiral dynamics from {AdS} space}}'',
  \href{http://dx.doi.org/10.1103/PhysRevD.70.046002}{{\em Phys. Rev.}
  {\bfseries D70} (2004) 046002},
\href{http://arxiv.org/abs/hep-th/0403279}{{\ttfamily arXiv:hep-th/0403279
  [hep-th]}}.

\bibitem{Gilbert:1964iy}
W.~Gilbert, ``{{Broken Symmetries and Massless Particles}}'',
\href{http://dx.doi.org/10.1103/PhysRevLett.12.713}{{\em Phys. Rev. Lett.}
  {\bfseries 12} (1964) 713--714}.

\bibitem{Schafer:2001bq}
T.~Schäfer, D.~T. Son, M.~A. Stephanov, D.~Toublan, and J.~J.~M. Verbaarschot,
  ``{{Kaon condensation and {Goldstone's} theorem}}'',
  \href{http://dx.doi.org/10.1016/S0370-2693(01)01265-5}{{\em Phys. Lett.}
  {\bfseries B522} (2001) 67--75},
\href{http://arxiv.org/abs/hep-ph/0108210}{{\ttfamily arXiv:hep-ph/0108210
  [hep-ph]}}.

\bibitem{Leutwyler:1993gf}
H.~Leutwyler, ``{{Nonrelativistic effective Lagrangians}}'',
  \href{http://dx.doi.org/10.1103/PhysRevD.49.3033}{{\em Phys. Rev.} {\bfseries
  D49} (1994) 3033--3043},
\href{http://arxiv.org/abs/hep-ph/9311264}{{\ttfamily arXiv:hep-ph/9311264
  [hep-ph]}}.

\bibitem{Hidaka:2012ym}
Y.~Hidaka, ``{{Counting rule for Nambu-Goldstone modes in nonrelativistic
  systems}}'', \href{http://dx.doi.org/10.1103/PhysRevLett.110.091601}{{\em
  Phys. Rev. Lett.} {\bfseries 110} no.~9, (2013) 091601},
\href{http://arxiv.org/abs/1203.1494}{{\ttfamily arXiv:1203.1494 [hep-th]}}.

\bibitem{Watanabe:2014fva}
H.~Watanabe and H.~Murayama, ``{{Effective Lagrangian for Nonrelativistic
  Systems}}'', \href{http://dx.doi.org/10.1103/PhysRevX.4.031057}{{\em Phys.
  Rev.} {\bfseries X4} no.~3, (2014) 031057},
\href{http://arxiv.org/abs/1402.7066}{{\ttfamily arXiv:1402.7066 [hep-th]}}.

\bibitem{Kaplan:2016}
J.~Kaplan, ``{{Lectures on AdS/CFT from the Bottom Up}}'',.
  \url{http://sites.krieger.jhu.edu/jared-kaplan/files/2016/05/AdSCFTCourseNotesCurrentPublic.pdf}.

\bibitem{Klebanov:2000me}
I.~R. Klebanov, ``{{TASI lectures: Introduction to the AdS/CFT
  correspondence}}'', in {\em {Strings, branes and gravity. Proceedings,
  Theoretical Advanced Study Institute, TASI'99, Boulder, USA, May 31-June 25,
  1999}}, pp.~615--650.
\newblock 2000.
\newblock
\href{http://arxiv.org/abs/hep-th/0009139}{{\ttfamily arXiv:hep-th/0009139
  [hep-th]}}.
\newblock

\bibitem{DHoker:2002nbb}
E.~D'Hoker and D.~Z. Freedman, ``{{Supersymmetric gauge theories and the AdS /
  CFT correspondence}}'', in {\em {Strings, Branes and Extra Dimensions: TASI
  2001: Proceedings}}, pp.~3--158.
\newblock 2002.
\newblock
\href{http://arxiv.org/abs/hep-th/0201253}{{\ttfamily arXiv:hep-th/0201253
  [hep-th]}}.
\newblock

\bibitem{Maldacena:2003nj}
J.~M. Maldacena, ``{{TASI 2003 lectures on AdS/CFT}}'', in {\em {Progress in
  string theory. Proceedings, Summer School, TASI 2003, Boulder, USA, June
  2-27, 2003}}, pp.~155--203.
\newblock 2003.
\newblock
\href{http://arxiv.org/abs/hep-th/0309246}{{\ttfamily arXiv:hep-th/0309246
  [hep-th]}}.
\newblock

\bibitem{Nastase:2007kj}
H.~Nastase, ``{{Introduction to AdS-CFT}}'',
\href{http://arxiv.org/abs/0712.0689}{{\ttfamily arXiv:0712.0689 [hep-th]}}.

\bibitem{Ramallo:2013bua}
A.~V. Ramallo, ``{{Introduction to the AdS/CFT correspondence}}'',
  \href{http://dx.doi.org/10.1007/978-3-319-12238-0_10}{{\em Springer Proc.
  Phys.} {\bfseries 161} (2015) 411--474},
\href{http://arxiv.org/abs/1310.4319}{{\ttfamily arXiv:1310.4319 [hep-th]}}.

\bibitem{Penedones:2016voo}
J.~Penedones, \href{http://dx.doi.org/10.1142/9789813149441_0002}{``{{TASI
  lectures on AdS/CFT}}'',} in {\em {Proceedings, Theoretical Advanced Study
  Institute in Elementary Particle Physics: New Frontiers in Fields and Strings
  (TASI 2015): Boulder, CO, USA, June 1-26, 2015}}, pp.~75--136.
\newblock 2017.
\newblock
\href{http://arxiv.org/abs/1608.04948}{{\ttfamily arXiv:1608.04948 [hep-th]}}.
\newblock

\bibitem{tHooft:1973alw}
G.~'t~Hooft, ``{A Planar Diagram Theory for Strong Interactions}'',
\href{http://dx.doi.org/10.1016/0550-3213(74)90154-0}{{\em Nucl. Phys.}
  {\bfseries B72} (1974) 461}.

\bibitem{Aharony:1999ti}
O.~Aharony, S.~S. Gubser, J.~M. Maldacena, H.~Ooguri, and Y.~Oz, ``{{Large N
  field theories, string theory and gravity}}'',
  \href{http://dx.doi.org/10.1016/S0370-1573(99)00083-6}{{\em Phys. Rept.}
  {\bfseries 323} (2000) 183--386},
\href{http://arxiv.org/abs/hep-th/9905111}{{\ttfamily arXiv:hep-th/9905111
  [hep-th]}}.

\bibitem{Henningson:1998gx}
M.~Henningson and K.~Skenderis, ``{{The Holographic Weyl anomaly}}'',
  \href{http://dx.doi.org/10.1088/1126-6708/1998/07/023}{{\em JHEP} {\bfseries
  07} (1998) 023},
\href{http://arxiv.org/abs/hep-th/9806087}{{\ttfamily arXiv:hep-th/9806087
  [hep-th]}}.

\bibitem{Henningson:1998ey}
M.~Henningson and K.~Skenderis, ``{{Holography and the Weyl anomaly}}'',
  \href{http://dx.doi.org/10.1002/(SICI)1521-3978(20001)48:1/3<125::AID-PROP125>3.0.CO;2-B,
  10.1002/(SICI)1521-3978(20001)48:1/3<125::AID-PROP125>3.3.CO;2-2}{{\em
  Fortsch. Phys.} {\bfseries 48} (2000) 125--128},
\href{http://arxiv.org/abs/hep-th/9812032}{{\ttfamily arXiv:hep-th/9812032
  [hep-th]}}.

\bibitem{Skenderis:2002wp}
K.~Skenderis, ``{{Lecture notes on holographic renormalization}}'',
  \href{http://dx.doi.org/10.1088/0264-9381/19/22/306}{{\em Class. Quant.
  Grav.} {\bfseries 19} (2002) 5849--5876},
\href{http://arxiv.org/abs/hep-th/0209067}{{\ttfamily arXiv:hep-th/0209067
  [hep-th]}}.

\bibitem{Skenderis:2008dh}
K.~Skenderis and B.~C. van Rees, ``{{Real-time gauge/gravity duality}}'',
  \href{http://dx.doi.org/10.1103/PhysRevLett.101.081601}{{\em Phys. Rev.
  Lett.} {\bfseries 101} (2008) 081601},
\href{http://arxiv.org/abs/0805.0150}{{\ttfamily arXiv:0805.0150 [hep-th]}}.

\bibitem{Skenderis:2008dg}
K.~Skenderis and B.~C. van Rees, ``{Real-time gauge/gravity duality:
  {Prescription, Renormalization and Examples}}'',
  \href{http://dx.doi.org/10.1088/1126-6708/2009/05/085}{{\em JHEP} {\bfseries
  05} (2009) 085},
\href{http://arxiv.org/abs/0812.2909}{{\ttfamily arXiv:0812.2909 [hep-th]}}.

\bibitem{Klebanov:1999tb}
I.~R. Klebanov and E.~Witten, ``{{AdS/CFT} correspondence and symmetry
  breaking}'', \href{http://dx.doi.org/10.1016/S0550-3213(99)00387-9}{{\em
  Nucl. Phys.} {\bfseries B556} (1999) 89--114},
\href{http://arxiv.org/abs/hep-th/9905104}{{\ttfamily arXiv:hep-th/9905104
  [hep-th]}}.

\bibitem{Breitenlohner:1982jf}
P.~Breitenlohner and D.~Z. Freedman, ``{{Stability in Gauged Extended
  Supergravity}}'',
\href{http://dx.doi.org/10.1016/0003-4916(82)90116-6}{{\em Annals Phys.}
  {\bfseries 144} (1982) 249}.

\bibitem{Mack:1975je}
G.~Mack, ``{{All Unitary Ray Representations of the Conformal Group SU(2,2)
  with Positive Energy}}'',
\href{http://dx.doi.org/10.1007/BF01613145}{{\em Commun. Math. Phys.}
  {\bfseries 55} (1977) 1}.

\bibitem{Minces:1999eg}
P.~Minces and V.~O. Rivelles, ``{{Scalar field theory in the {AdS/CFT}
  correspondence revisited}}'',
  \href{http://dx.doi.org/10.1016/S0550-3213(99)00833-0}{{\em Nucl. Phys.}
  {\bfseries B572} (2000) 651--669},
\href{http://arxiv.org/abs/hep-th/9907079}{{\ttfamily arXiv:hep-th/9907079
  [hep-th]}}.

\bibitem{Minces:2001zy}
P.~Minces and V.~O. Rivelles, ``{{Energy and the AdS / CFT correspondence}}'',
  \href{http://dx.doi.org/10.1088/1126-6708/2001/12/010}{{\em JHEP} {\bfseries
  12} (2001) 010},
\href{http://arxiv.org/abs/hep-th/0110189}{{\ttfamily arXiv:hep-th/0110189
  [hep-th]}}.

\bibitem{Rivelles:2003ge}
V.~O. Rivelles, ``{{Quantization in AdS and the AdS / CFT correspondence}}'',
  \href{http://dx.doi.org/10.1142/S0217751X03015544}{{\em Int. J. Mod. Phys.}
  {\bfseries A18} (2003) 2099--2108},
\href{http://arxiv.org/abs/hep-th/0301025}{{\ttfamily arXiv:hep-th/0301025
  [hep-th]}}.

\bibitem{Witten:2001ua}
E.~Witten, ``{{Multitrace operators, boundary conditions, and AdS / CFT
  correspondence}}'',
\href{http://arxiv.org/abs/hep-th/0112258}{{\ttfamily arXiv:hep-th/0112258
  [hep-th]}}.

\bibitem{Marolf:2006nd}
D.~Marolf and S.~F. Ross, ``{{Boundary Conditions and New Dualities: Vector
  Fields in {AdS/CFT}}}'',
  \href{http://dx.doi.org/10.1088/1126-6708/2006/11/085}{{\em JHEP} {\bfseries
  11} (2006) 085},
\href{http://arxiv.org/abs/hep-th/0606113}{{\ttfamily arXiv:hep-th/0606113
  [hep-th]}}.

\bibitem{Bzowski:2016kni}
A.~Bzowski, ``{{Dimensional renormalization in AdS/CFT}}'',
\href{http://arxiv.org/abs/1612.03915}{{\ttfamily arXiv:1612.03915 [hep-th]}}.

\bibitem{Petkou:1999fv}
A.~Petkou and K.~Skenderis, ``{{A Nonrenormalization theorem for conformal
  anomalies}}'', \href{http://dx.doi.org/10.1016/S0550-3213(99)00514-3}{{\em
  Nucl. Phys.} {\bfseries B561} (1999) 100--116},
\href{http://arxiv.org/abs/hep-th/9906030}{{\ttfamily arXiv:hep-th/9906030
  [hep-th]}}.

\bibitem{Mir:2013pca}
M.~Mir, ``{{On Holographic Weyl Anomaly}}'',
  \href{http://dx.doi.org/10.1007/JHEP10(2013)084}{{\em JHEP} {\bfseries 10}
  (2013) 084},
\href{http://arxiv.org/abs/1307.5514}{{\ttfamily arXiv:1307.5514 [hep-th]}}.

\bibitem{Kraus:1999di}
P.~Kraus, F.~Larsen, and R.~Siebelink, ``{{The gravitational action in
  asymptotically AdS and flat space-times}}'',
  \href{http://dx.doi.org/10.1016/S0550-3213(99)00549-0}{{\em Nucl. Phys.}
  {\bfseries B563} (1999) 259--278},
\href{http://arxiv.org/abs/hep-th/9906127}{{\ttfamily arXiv:hep-th/9906127
  [hep-th]}}.

\bibitem{deBoer:1999tgo}
J.~de~Boer, E.~P. Verlinde, and H.~L. Verlinde, ``{{On the holographic
  renormalization group}}'',
  \href{http://dx.doi.org/10.1088/1126-6708/2000/08/003}{{\em JHEP} {\bfseries
  08} (2000) 003},
\href{http://arxiv.org/abs/hep-th/9912012}{{\ttfamily arXiv:hep-th/9912012
  [hep-th]}}.

\bibitem{deBoer:2000cz}
J.~de~Boer, ``{{The Holographic renormalization group}}'',
  \href{http://dx.doi.org/10.1002/1521-3978(200105)49:4/6<339::AID-PROP339>3.0.CO;2-A}{{\em
  Fortsch. Phys.} {\bfseries 49} (2001) 339--358},
\href{http://arxiv.org/abs/hep-th/0101026}{{\ttfamily arXiv:hep-th/0101026
  [hep-th]}}.

\bibitem{Martelli:2002sp}
D.~Martelli and W.~Mueck, ``{{Holographic renormalization and Ward identities
  with the Hamilton-Jacobi method}}'',
  \href{http://dx.doi.org/10.1016/S0550-3213(03)00060-9}{{\em Nucl. Phys.}
  {\bfseries B654} (2003) 248--276},
\href{http://arxiv.org/abs/hep-th/0205061}{{\ttfamily arXiv:hep-th/0205061
  [hep-th]}}.

\bibitem{Papadimitriou:2004ap}
I.~Papadimitriou and K.~Skenderis, ``{{AdS / CFT correspondence and
  geometry}}'', \href{http://dx.doi.org/10.4171/013-1/4}{{\em IRMA Lect. Math.
  Theor. Phys.} {\bfseries 8} (2005) 73--101},
\href{http://arxiv.org/abs/hep-th/0404176}{{\ttfamily arXiv:hep-th/0404176
  [hep-th]}}.

\bibitem{Papadimitriou:2004rz}
I.~Papadimitriou and K.~Skenderis, ``{{Correlation functions in holographic RG
  flows}}'', \href{http://dx.doi.org/10.1088/1126-6708/2004/10/075}{{\em JHEP}
  {\bfseries 10} (2004) 075},
\href{http://arxiv.org/abs/hep-th/0407071}{{\ttfamily arXiv:hep-th/0407071
  [hep-th]}}.

\bibitem{Papadimitriou:2016yit}
I.~Papadimitriou, ``{{Lectures on Holographic Renormalization}}'',
\href{http://dx.doi.org/10.1007/978-3-319-31352-8_4}{{\em Springer Proc. Phys.}
  {\bfseries 176} (2016) 131--181}.

\bibitem{Gubser:2008px}
S.~S. Gubser, ``{{Breaking an {Abelian} gauge symmetry near a black hole
  horizon}}'', \href{http://dx.doi.org/10.1103/PhysRevD.78.065034}{{\em Phys.
  Rev.} {\bfseries D78} (2008) 065034},
\href{http://arxiv.org/abs/0801.2977}{{\ttfamily arXiv:0801.2977 [hep-th]}}.

\bibitem{Hartnoll:2008vx}
S.~A. Hartnoll, C.~P. Herzog, and G.~T. Horowitz, ``{{Building a Holographic
  Superconductor}}'',
  \href{http://dx.doi.org/10.1103/PhysRevLett.101.031601}{{\em Phys. Rev.
  Lett.} {\bfseries 101} (2008) 031601},
\href{http://arxiv.org/abs/0803.3295}{{\ttfamily arXiv:0803.3295 [hep-th]}}.

\bibitem{Argurio:2015via}
R.~Argurio, A.~Marzolla, A.~Mezzalira, and D.~Naegels, ``{{Note on holographic
  nonrelativistic Goldstone bosons}}'',
  \href{http://dx.doi.org/10.1103/PhysRevD.92.066009}{{\em Phys. Rev.}
  {\bfseries D92} no.~6, (2015) 066009},
\href{http://arxiv.org/abs/1507.00211}{{\ttfamily arXiv:1507.00211 [hep-th]}}.

\bibitem{Balasubramanian:1998de}
V.~Balasubramanian, P.~Kraus, A.~E. Lawrence, and S.~P. Trivedi,
  ``{{Holographic probes of anti-de Sitter space-times}}'',
  \href{http://dx.doi.org/10.1103/PhysRevD.59.104021}{{\em Phys. Rev.}
  {\bfseries D59} (1999) 104021},
\href{http://arxiv.org/abs/hep-th/9808017}{{\ttfamily arXiv:hep-th/9808017
  [hep-th]}}.

\bibitem{Girardello:1999bd}
L.~Girardello, M.~Petrini, M.~Porrati, and A.~Zaffaroni, ``{{The Supergravity
  dual of N=1 superYang-Mills theory}}'',
  \href{http://dx.doi.org/10.1016/S0550-3213(99)00764-6}{{\em Nucl. Phys.}
  {\bfseries B569} (2000) 451--469},
\href{http://arxiv.org/abs/hep-th/9909047}{{\ttfamily arXiv:hep-th/9909047
  [hep-th]}}.

\bibitem{Argurio:2014rja}
R.~Argurio, D.~Musso, and D.~Redigolo, ``{{Anatomy of new {SUSY} breaking
  holographic {RG} flows}}'',
  \href{http://dx.doi.org/10.1007/JHEP03(2015)086}{{\em JHEP} {\bfseries 03}
  (2015) 086},
\href{http://arxiv.org/abs/1411.2658}{{\ttfamily arXiv:1411.2658 [hep-th]}}.

\bibitem{Argurio:2017xxx}
R.~Argurio, J.~Hartong, A.~Marzolla, and D.~Naegels, ``{Symmetry breaking in
  holographic theories with Lifshitz scaling}'',
\href{http://arxiv.org/abs/1709.08383}{{\ttfamily arXiv:1709.08383 [hep-th]}}.

\end{thebibliography}\endgroup
\bibliographystyle{utphys}

	
\end{document}